\documentclass[a4paper,11pt]{article}
\usepackage{jcappub} 
\usepackage{verbatim}
\usepackage{multirow}
\usepackage{float}
\restylefloat{table}
\usepackage{pifont}
\usepackage{epsfig}
\usepackage{graphicx}
\usepackage{boldline}
\usepackage{booktabs}
\usepackage{multirow}
\usepackage{dcolumn}
\usepackage{amsmath}
\usepackage{mathtools}
\usepackage{amsfonts}
\usepackage{amssymb}
\usepackage{epstopdf}
\usepackage{multirow,bigdelim}
\usepackage{bm}
\usepackage{braket}
\usepackage{enumitem}
\usepackage[table]{xcolor}
\newcommand{\nc}{\newcommand}
\nc{\x}{{\bf{x}}}
\nc{\q}{{\bf{q}}}
\nc{\cF}{{\cal F}}
\nc{\cO}{{\cal O}}
\nc {\cG}{{\cal G}}
\definecolor{navyblue}{rgb}{0.0, 0.0, 0.5}
\definecolor{royalblue}{rgb}{0.25, 0.41, 0.88}
\definecolor{cadmiumgreen}{rgb}{0.0, 0.42, 0.24}
\definecolor{blue-violet}{rgb}{0.54, 0.17, 0.89}
\definecolor{darkviolet}{rgb}{0.58, 0.0, 0.83}
\definecolor{orange(colorwheel)}{rgb}{1.0, 0.5, 0.0}

\usepackage{hyperref}
\hypersetup{colorlinks=true,linkcolor=royalblue,citecolor=magenta,pdfencoding=auto}
\usepackage{pifont}
\usepackage{dcolumn}
\usepackage{colortbl}

\newcommand{\ba}{\begin{eqnarray}}
\newcommand{\ea}{\end{eqnarray}}
\newcommand{\ep}{\epsilon}
\newcommand*{\rom}[1]{\expandafter\@slowromancap\romannumeral #1@}
\newcommand\be{\begin{eqnarray}}
\newcommand\ee{\end{eqnarray}}

\newcommand{\nn}{\nonumber}
\newcommand{\bk}{{\bf k}}

\newcommand{\bq}{{\bf q}}
\newcommand{\bx}{{\bf x}}

\newcommand{\bfk}{{\bk}}
\newcommand{\bfq}{{\bq}
\newcommand{\bfx}{{\bx}}}

\newcommand{\bfx}{{\bf{x}}}

\def\bk{{\bf k}}

\def\bq{{\bf q}}
\def\bx{{\bf x}}

\def \bfk{{\bk}}
\def \bfq{{\bq}
\def \bfx{{\bx}}}

\def \ep {\epsilon}

\renewcommand{\O}{\mathcal{O}}

\def \ep {\epsilon}

\def\bk{{\bf k}}

\def\bq{{\bf q}}
\def\bx{{\bf x}}

\def \bfk{{\bk}}
\def \bfq{{\bq}
	\def \bfx{{\bx}}}

\def\bk{{\bf k}}

\def\bq{{\bf q}}
\def \vpi{\varphi}
\def\bx{{\bf x}}

\def \bfk{{\bk}}
\def \bfq{{\bq}
	\def \bfx{{\bx}}}

\def \ep {\epsilon}
\def \cI {{\cal I}}
\def \cB {{\cal B}}
\def \la {\langle}
\def \ra {\rangle}

\newcommand{\e}{\epsilon}

\definecolor{magenta(process)}{rgb}{1.0, 0.0, 0.56}

\definecolor{darkspringgreen}{rgb}{0.09, 0.45, 0.27}

\definecolor{royalblue(web)}{rgb}{0.25, 0.41, 0.88}

\title{\centering Solid Soft Theorems}

\author[a]{Enrico Pajer,}
\author[a]{Sadra Jazayeri,}
\author[b]{Drian van der Woude,}

\affiliation[a]{Department of Applied Mathematics and Theoretical Physics, Centre for Mathematical Sciences,
University of Cambridge, Wilberforce Road, Cambridge CB3 0WA, UK}
\affiliation[b]{Institute for Theoretical Physics and Center for Extreme Matter and Emergent Phenomena,
	Utrecht University, 
	Princetonplein 5, 3584 CC Utrecht, The Netherlands}
\emailAdd{ep551@cam.ac.uk}
\emailAdd{sj571@cam.ac.uk}
\emailAdd{drianvanderwoude@gmail.com}

\abstract{\noindent  
We derive cosmological soft theorems for solids coupled to gravity. To this end, we first derive all cosmological adiabatic modes for solids, which display the interesting novelty of non-vanishing anisotropic stresses on large scales. Then, from the corresponding symmetries of the action of perturbations we compute the leading order related soft theorems using the operator product expansion. For the scalar bispectrum, we re-derive the result that Maldacena's consistency relation is recovered only upon angular averaging over the long mode direction. In addition, we find theorems for soft tensor and vector perturbations. In passing, we also clarify the derivation of these soft theorems in gauges where no residual diffeomorphisms exist.
}

\begin{document}
\maketitle
\flushbottom

\vspace{1cm}

\section{Introduction}

Correlation functions for cosmological perturbations are the primary observables to constrain cosmological parameters and inflationary physics. Particularly useful in this respect are model independent results for correlators that can be established for large classes of models. A prominent example is Maldacena's consistency relation \cite{Maldacena:2002vr}: in single-clock inflation \cite{Creminelli:2004yq}, the limit in which one of the momenta in an $  n $-point function becomes much smaller than the others is fixed by lower $  n $-point functions. For example, the squeezed primordial bispectrum is fixed in terms of the primordial power spectrum. As a consequence, the bispectra of cosmological observables such as the anisotropies and spectral distortions of the Cosmic Microwave Background (CMB) and Large Scale Structures (LSS) in the squeezed limit are determined by the late time evolution and receive vanishing contributions from inflationary physics (see e.g.~\cite{Pajer:2013ana}). Maldacena's result is similar to soft theorems in particle physics and has been understood to arise as consequence of non-linearly realized symmetries \cite{Hinterbichler:2012nm,Hinterbichler:2013dpa,Assassi:2012zq,Creminelli:2011rh,Creminelli:2012ed,Creminelli:2013cga,Bordin:2016ruc} induced by adiabatic modes \cite{Weinberg:2003sw,Hinterbichler:2013dpa,Mirbabayi:2014hda,Pajer:2017hmb,Finelli:2017fml}. Adiabatic modes are physical perturbations that are locally indistinguishable from change of coordinates and therefore can be traded for diffeomorphism in correlators.

Two important loopholes have been identified in consistency relations. The first is the assumption that curvature perturbations $ \zeta  $ freeze on superHubble scales. In non-attractor models such as Ultra-Slow-Roll (USR) inflation \cite{Kinney:2005vj} this does not happen and $  \zeta $ continues to evolve in time even for $  q\ll aH $. The bispectrum in these models has been shown to violate Maldacena's consistency relation by a large amount \cite{Namjoo:2012aa,Chen:2013aj}. The violation results from the fact that the time dependence of the long mode does not match that of the adiabatic mode (see e.g.~\cite{Akhshik:2015rwa,Mooij:2015yka}). A more basic relation can still be established even for USR \cite{Hui:2018cag}, but it does not lead to a relation that can be verified observationally. In addition, some USR models are symmetric under a constant shift of the scalar field. In this case, new soft theorems were derived in \cite{Finelli:2017fml,Bravo:2017wyw,Finelli:2018upr}, using methods very similar to those we will employ in this work. For the transition from USR to slow-roll inflation, see the recent discussions in \cite{Cai:2017bxr,Passaglia:2018ixg}.

The second loophole is the possibility of a non-standard symmetry-breaking pattern, as realized for example in solid inflation \cite{Gruzinov:2004ty,Endlich:2012pz} (see also \cite{Bartolo:2015qvr,Ricciardone:2016lym,Piazza:2017bsd,Fujita:2018ehq} for other possibilities). Naively, this model contains three scalar fields. But on the relevant solid background, spatial as well as internal rotations and translations are spontaneously broken in such a way that a diagonal combination survives, just as in the spontaneous symmetry probing mechanism of \cite{Nicolis:2011pv}. As a result, the three would-be scalars reorganize themselves on this background into one scalar and one transverse vector. Even though there is a single scalar perturbation, the bispectrum violates Maldacena's consistency relation \cite{Endlich:2012pz,Endlich:2013jia,Akhshik:2014bla}. The violation can be traced back to the fact that a generic long wavelength curvature perturbation with $  q\rightarrow 0 $ is still locally measurable and therefore not an adiabatic mode. This is in contrast with standard inflation in which such a perturbations is locally (i.e.~up to order $  q^{2} $) indistinguishable from a change of coordinates. Remarkably, the authors of \cite{Bordin:2017ozj} found that even though the full squeezed bispectrum is unconstrained, its angular average still obeys Maldacena's consistency relation. They explain their finding by showing that a specific \textit{isotropic} long-wavelength curvature perturbation is still a local change of coordinates. 


\paragraph{Summary of results} In this work we will understand this and other soft theorems as result of the existence of generalized adiabatic modes for solids. These are physical perturbations that are locally indistinguishable from a change of coordinates and an internal symmetry transformation. At the technical level, this is the first time that adiabatic modes are studied for cosmological solutions that support anisotropic stresses at superHubble scales, as it is the case in the solid. At the phenomenological level, we find model-independent relations among correlators that fix some of the leading coefficients in the squeezed limit of any $ n$-point correlator. We mostly focus on bispectra and trispectra, but similar soft theorems hold for arbitrary correlators. To be more precise, let us introduce the following parameterization of the soft scalar, tensor and vector bispectra:
\begin{align}
\label{Legendreexp}
\la \zeta_\bfq \zeta_{\bfk-\frac{1}{2}\bfq}\zeta_{-\bfk-\frac{1}{2}\bfq}\ra &=P_\zeta(q)P_\zeta(k)   \sum_{n,l} \left( \frac{q}{k} \right)^{n}a_l^{(n)}(k)P_l(\hat q \cdot \hat k) \,,\\ \nn
\la \gamma^s_{\bfq}\zeta_{\bfk-\frac{1}{2}\bfq}\zeta_{-\bfk-\frac{1}{2}\bfq}\ra &= P_\gamma(q)P_\zeta(k) \ep^s_{ij}(\hat{q})\hat{k}^i\hat{k}^j   \sum_{n,l} \left( \frac{q}{k} \right)^{n}b_l^{(n)}(k)P_l(\hat q \cdot \hat k)\,,\\
\la  \vpi^s_{V\bfq}\zeta_{\bfk-\frac{1}{2}\bfq}\zeta_{-\bfk-\frac{1}{2}\bfq}\ra &=P_V(q)P_\zeta(k)\ep^s_{i}(\hat{q})\hat{k}^i  \sum_{n,l} \left( \frac{q}{k} \right)^{n}c_l^{(n)}(k)P_l(\hat q \cdot \hat k)\,.
\end{align}
Here we have first expanded in $  q\rightarrow 0 $ assuming a regular analytic behavior (of the related OPE's) and then rewritten each term in a basis of Legendre polynomials $P_{\ell}$ (not to be confused with the power spectrum $P_{\zeta} (k)$). Also, $\ep^s_{ij}$ is the polarization tensor for the soft helicity-$s $ graviton, and $\ep^s_i$ denotes the polarization vector of a soft helicity-$s  $ transverse phonon. Finally, in the coefficients $  a $, $  b $ and $  c $, the upper labels between parenthesis indicates the order in $  q $, while the lower labels refer to the order of the associated Legendre polynomial.

In order to match the well defined parity under $  \vec k \rightarrow -\vec k $ of the left-hand side correlators (see e.g.~appendix A of \cite{Assassi:2015jqa}) for any $  n,l\in \mathbb{N} $ we can already conclude that 
\begin{align}\label{parity}
a^{(n)}_{2l+1}=b^{(n)}_{2l+1}=c^{(n)}_{2l}=0\,.
\end{align} 
Also, just from the structure of the Operator Product Expansion (OPE) we see that (see Sec.~\ref{sec3}) 
\begin{align}
a^{(n)}_{2l}&=0 \quad \text{for} \quad   2l\geq n+2\,, \\
b^{(n)}_{2l}&=0 \quad \text{for} \quad   2l\geq n\,,\\
c^{(n)}_{2l+1}&=0 \quad \text{for} \quad   2l+1\geq n\,.
\end{align}  
We are therefore left with 
\begin{align}
\label{Legendreexp}
\la \zeta_\bfq \zeta_{\bfk-\frac{1}{2}\bfq}\zeta_{-\bfk-\frac{1}{2}\bfq}\ra &=P_\zeta(q)P_\zeta(k)  \left[ a_0^{(0)}(k)+a_2^{(0)}(k)P_{2}(\cos\theta) +{\cal O}(q^2)  \right]\,, \\ \nn
\la \gamma^s_{\bfq}\zeta_{\bfk-\frac{1}{2}\bfq}\zeta_{-\bfk-\frac{1}{2}\bfq}\ra &= P_\gamma(q)P_\zeta(k) \ep^s_{ij}(\hat{q})\hat{k}^i\hat{k}^j  \left[  b^{(0)}_0(k)+{\cal O}(q^2)  \right]\,,\\ \nn
\la  \vpi^s_{V\bfq}\zeta_{\bfk-\frac{1}{2}\bfq}\zeta_{-\bfk-\frac{1}{2}\bfq}\ra &=i\,P_V(q)P_\zeta(k)\ep^s_{i}(\hat{q})\hat{k}^i  \left[ q\,c_1^{(1)}(k)P_1(\cos\theta)+{\cal O}(q^3)  \right]\,,
\end{align}
In Sec.~\ref{sec3} we will prove that these coefficients satisfy the following consistency relations as consequence of the Ward-Takahashi identities for the solid adiabatic modes:
\begin{align}
a_{0}^{(0)}&=(n_s-1)\,,\label{malda}\\
 2b^{(0)}_0(k)+\dfrac{1}{2}a_2^{(0)}(k)&=3+(1-n_{s})\,, \label{fre}\\
2b^{(0)}_{0}(k)+\,c_{1}^{(1)}(k)&=3+(1-n_{s})\,.
\end{align}
The relation \eqref{malda} is the angular averaged Maldacena's consistency relation and was already derived in \cite{Bordin:2017ozj}. Also, the relation \eqref{fre} was previously noted in \cite{Bordin:2016ruc}.
Infinitely many relations to higher order in $ q$ can also be derived by the same methods. We illustrate this point by deriving an ${\cal O}(q)$ soft theorem for a sample trispectrum in Sec.~\ref{sample}. All the soft theorems we derived in this work are valid for any accelerated FLRW spacetime, assuming a Bunch-Davies vacuum (i.e. we do not assume any internal dilation symmetry of the solid). In particular, $  (n_{s}-1) $ in the above relations does not need to be small.

Two final comments are in order. First, we emphasize that these relations can in principle be confirmed by observations as they involve only quantities that are theoretically accessible to late time cosmological observations, such as the power spectrum and bispectrum. This is in contrast with other soft theorems that involve time derivative of correlators, as for example in shift-symmetric cosmologies \cite{Finelli:2017fml}. Second, assuming that reheating in solid inflation does not alter the predictions substantially, Planck data found the bound\footnote{Actually the Planck bound assumes that the non-Gaussian shape is as in \eqref{Legendreexp} for all configurations, while for the solid this is only valid in the squeezed limit. But since most of the signal is in the squeezed limit, this should not make much of a difference.} \cite{Ade:2015ava}
\begin{align}
a_{2}^{(0)}&=-16 f_{NL}^{L=2}=20.8 \pm 62.4 \,,
\end{align} 
for the KSW estimator and the SMICA map at $68\%$ CL. The relation \eqref{fre} then implies the bound 
\begin{align}
b_{0}^{(0)}&=-3.7\pm15.6 & \text{(indirect from Planck \cite{Ade:2015ava})}\,.
\end{align} 
This is much tighter than the bounds that can be derived from the BTT bispectrum of the CMB \cite{Meerburg:2016ecv,Bordin:2016ruc} and from the anisotropies in the scalar power spectrum due to the inefficient erasing of anisotropies during inflation \cite{Bartolo:2013msa,Akhshik:2014gja,Bordin:2016ruc}, which instead give respectively
\begin{align}
|b_{0}^{(0)}|&\lesssim 290 \sqrt{\frac{0.07}{r}} &\text{(BTT)}\,, \\
|b_{0}^{(0)}|&\lesssim 600 \sqrt{\frac{0.07}{r}} &(P_{\zeta}\text{ ansisotropy})\,.
\end{align}


\section{Solid Adiabatic Modes}

In this section, after reviewing cosmological solutions for solids along the lines of \cite{Endlich:2012pz}, we derive an infinite number of new adiabatic modes present in this setup. These adiabatic modes are very distinct from those present for perfect fluids (see e.g.~\cite{Pajer:2017hmb} for a recent comprehensive discussion) and are related to the symmetries of the theory. To clarify their gauge dependence, we discuss adiabatic modes in two different gauges.


\subsection{Cosmological solution with solids: a review}

Following \cite{Leutwyler:1996er,Dubovsky:2005xd, Endlich:2010hf,Endlich:2012pz,Nicolis:2015sra}, a solid state of matter can be defined by the following symmetry breaking pattern
\be\label{symmpattern}
\text{ISO(3)}_{\text{internal}}\times \text{ISO(3)}_{\text{space}}\to \text{ISO(3)}_{\text{diagonal}}\,.
\ee
Here, in light of cosmological application, we restricted ourselves to solids that preserve isotropy, sometimes called ``jellies''. 
ISO(3)$  _{\text{internal}} $ is defined to act on three scalar fields $  \phi^{I} $, $  I=1,2,3 $, according to
\be
\phi^I\to O^I_J\,\phi^J+c^J\,, \quad\quad \forall \,(O,c)\in \text{ISO(3)}_{\text{internal}}\,,
\ee
while the action of ISO(3)$  _{\text{space}} $ is given as always by
\be
\phi^I(t,x^i)\to \phi^I(t,\tilde{O}^i_j\,x^j+\tilde{c}^j)\,, \quad\quad \forall \,(\tilde{O},\tilde{c})\in \text{ISO(3)}_{\text{space}}\,.
\ee
In the ground state of the solid, both copies of ISO(3) are broken by the scalar field expectation values  
\be\label{gs}
\langle \phi^I\rangle=x^I\,.
\ee
However, a diagonal combination of the two ISO(3) groups leaves this ground state unchanged and is therefore left unbroken, namely
\be
x^i\to O^i_j\,x^j+c^i\,, \quad\quad\quad \phi^i\to \Big(O^{-1}\Big)^i_j\phi^j-c^j\,.
\ee
This defines the ISO(3)$ _{\text{diagonal}}  $ group, which furnishes a scalar-vector-tensor decomposition for the metric and the energy-momentum tensor perturbations. Under the action of this unbroken symmetry group, field and space indices transform in the same way, thus we refer to both of them by $i,j,...$, hereafter. This unbroken symmetry is responsible for the statistical homogeneity and isotropy of all correlators, as needed for cosmological applications.

The ISO(3)$  _{\text{internal}} $ symmetry enables us to write down the most generic low-energy theory for these three scalar fields, which we will consider only at lowest order in spacetime derivatives. Minimally coupling this theory to gravity results in \cite{Endlich:2012pz}
\be
S=\int d^4x \sqrt{-g}\left[\dfrac{1}{2}M_p^2 R+\cF(X,Y,Z)\right]\,,
\ee	
where $  \cF $ is an arbitrary function,
\ba
\cB_{IJ}\equiv g^{\mu\nu}\partial_{\mu}\phi^I\partial_{\nu}\phi^J\,,\quad X\equiv[\cB]\,,\quad Y\equiv\dfrac{[\cB^2]}{[\cB]^2}\,,\quad Z\equiv\dfrac{[\cB^3]}{[\cB]^3}\,,
\ea
and $  \left[  \dots\right] $ indicates taking the trace over $  I,J $ indices. The above action admits FLRW solutions with
\be
\langle\phi^I \rangle=x^I\,,
\ee
and an isotropic energy momentum tensor, parameterized by the following energy density and pressure
\ba
\bar{\rho}&=&-\cF(\bar  X(t),\bar Y,\bar  Z)\,, \quad\quad\bar{p}=\cF(\bar  X(t),\bar  Y,\bar  Z)-\dfrac{2}{a^2}\cF_X(\bar  X(t),\bar Y,\bar  Z)\,,\\ \nn
\bar X(t)&=&\dfrac{3}{a^2}\,,\hspace{3cm} \bar Y=1/3\,,\quad \quad \, \bar Z=1/9\,.
\ea
As thoroughly investigated in \cite{Endlich:2012pz}, letting $\cF$ depend on $X$ weakly leads to a successful period of inflation. Nevertheless, in the remainder of this paper, we do not assume that $  \cF_{X} $ is small, nor do we assume any specific time dependence for $a(t)$. \textit{Our results remain valid for any FLRW background, whether or not it is close to a de Sitter spacetime.}

It is useful to have the fully non-linear energy-momentum tensor, which is given by
\ba\nonumber
T_{\mu\nu}=\cF g_{\mu\nu}-2 \partial_\mu \phi^I \partial_{\nu}\phi^J\left[ \left(\cF_X-\dfrac{2\cF_Y Y}{X}-\dfrac{3\cF_Z Z}{X} \right)\delta^{IJ}+\dfrac{2\cF_Y}{X^2}\cB^{IJ}+\dfrac{3\cF_Z \cB_{IK}{\cB_{JK}}}{X^3}\right]\,,
\ea
where $  \cF_{X} $, $ \cF_{Y} $ and  $  \cF_{Z} $  indicate partial derivatives of $  \cF $. 

Let us turn to investigate linear perturbation theory around the aforementioned homogeneous background. 
We follow the notations in \cite{Weinberg:2008zzc}\footnote{With one exception that we use $\gamma_{ij} $ in place of $D_{ij}$ for tensor perturbations.} for first order perturbations in the metric and the matter fields. We denote the components of the linearly perturbed metric by
\ba
\label{mconv}
ds^2=-(1+E)dt^2+2a (\partial_i F+G_i)dt dx^i+a^2 \left[ (1+A)\delta_{ij}+\partial_{i}\partial_{j}B+2\partial_{(i} C_{j)}+\gamma_{ij} \right]\,,
\ea
while for the matter perturbation we have
\be
\label{thermo}
\delta T_{00}&=&- \bar \rho h_{00}+\delta \rho\,, \\
\delta T_{0i}&=& \bar  p h_{0i}-(\rho+p)\delta u_i\,,\quad\quad \delta u_i=\partial_i \delta u+\delta u_i^V\,,\\
\delta T_{ij}&=& \bar p h_{ij}+a^2\left(\delta p\,\delta_{ij}+\partial_i \partial_j \pi+\partial_i \pi_j+\partial_j \pi_i+\pi_{ij}\right)\,.
\ee
Here, $\lbrace E,F,A, B, \delta \rho, \delta p, \pi\rbrace$ transform as scalars under the action of the unbroken $SO(3)$ symmetry group. Moreover $\lbrace G_i,C_i, \pi_i\rbrace$ transform as vectors and satisfy
\be
\partial_i C_i=\partial_i G_i=\partial_i \pi_i=0\,.
\ee
Finally, $\lbrace\gamma_{ij},\pi_{ij}\rbrace$ stand for helicity-two excitations, a.k.a. transverse traceless ``tensors'', 
\be
\gamma_{ii}=\partial_i \gamma_{ij}=\pi_{ii}=\partial_i \pi_{ij}=0.
\ee
We define the matter sector perturbations (dubbed ``phonons") through 
\be
\phi^I \equiv x^I+\vpi^I\,.
\ee
Phonons may be SVT-decomposed further into a longitudinal and a transverse component 
\be
\vpi^i \equiv \partial_i \vpi_L+\vpi^i_V\,.
\ee
Quantities defined in \eqref{thermo} can be related to the solid displacement fields and metric perturbations by 
\ba
\label{deltrho}
\delta \rho&=&-\cF_X \delta X\,,\\
\label{deltau}
\delta u&=&-a^2\dot{\vpi}_L+a \, F\,,\\
\label{deltauV}
\delta u^V_i&=&-a^2 \dot{\vpi}^V_i+a\,G_i\,,\\
\label{deltap}
\delta p&=&\left(\dfrac{2\cF_X}{a^2}+\dfrac{4(\cF_Y+\cF_Z)}{X^2a^4}\right)A+\\ \nn
&&\hspace{2cm}+\left(\cF_X-\dfrac{2}{a^2}\cF_{XX}+\dfrac{4}{3a^2X^2}(\cF_Y+\cF_Z)\right) \delta X\,,\\
\label{pij}
\pi^S&=&\left(\dfrac{2\cF_X}{a^2}+\dfrac{4}{9}(\cF_Y+\cF_Z)\right)\left(B-2\vpi_L\right)\,,\\
\label{piV}
\pi^V_i&=&\left(\dfrac{2\cF_X}{a^2}+\dfrac{4}{9}(\cF_Y+\cF_Z)\right)\left(C_i-\vpi_i^V\right)\,,\\ 
\label{pijT}
\pi_{ij}&=&\left(\dfrac{2\cF_X}{a^2}+\dfrac{4}{9}(\cF_Y+\cF_Z)\right)\,\gamma_{ij}\,,
\ea
where
\be
\label{deltaX}
\delta X=-\dfrac{1}{a^2}\left(3A+\nabla^2 B\right)+\dfrac{2}{a^2}\nabla^2 \vpi_L\,.
\ee
Notice that all these equations hold in an arbitrary gauge. \\

In the next subsection, we will derive the time dependence of the adiabatic modes solely by symmetry principles. However, one can also derive the dynamical equations for the scalar, vector and tensor perturbations by explicitly solving the linearized Einstein equations. We do this in the following. For tensor perturbations, one finds
\be
\ddot{\gamma}_{ij}+3H\dot{\gamma}_{ij}+\dfrac{k^2}{a^2} \gamma_{ij}=\dfrac{2}{M_p^2}\pi_{ij}^T\,.
\ee
Plugging in \eqref{pijT}, and letting $k\to 0$ yields
\be
\ddot{\gamma}_{ij}+3H\dot{\gamma}_{ij}-\left[\dfrac{2\cF_X}{a^2}+\dfrac{4}{9}(\cF_Y+\cF_Z)\right]\gamma_{ij}=0\,.
\ee
As emphasized in \cite{Endlich:2012pz}, the graviton acquires a mass that can be expressed in terms of $\dot{H}$ and the sound speed of transverse phonons $c_T$, i.e.
\be
m_{\gamma}^2 \equiv \left[\dfrac{2\cF_X}{a^2}+\dfrac{4}{9}(\cF_Y+\cF_Z)\right]=4\dot{H}c_T^2\,.
\ee
We do not explicitly write down the equations of motion for scalars and vectors, however, the time dependence of them in the zero momentum limit obeys the same ODE as above, 
\be
\label{cI}
\ddot{\cI}+3H\dot{\cI}-4\dot{H}c_T^2\,\cI=0\,.
\ee
In the remainder of this paper we use $\cI_a(t)$, with $  a=1,2 $, to denote two independent solutions of this equation.

\subsection{Large gauge transformations}

Adiabatic modes are related to the invariance of the action under an infinitesimal coordinate transformation 
\be
x^\mu\to x^\mu+\ep^\mu(x)\,.
\ee
The diffeomorphisms that give rise to non-trivial charge operators are ``large'' gauge transformations, meaning that they do not fall off at spatial infinity\footnote{It must be beard in mind that "large" gauge transformation can still be infinitesimal---they do not decay at infinity, nevertheless they are multiplied by infinitesimal parameters of the Lie group.}. Conversely, ``small'' gauge transformations, which decay at large distances----and hence do not contribute to any Noether charge---will be fixed by a (local) gauge choice.

For the metric perturbations $h_{\mu\nu}$ defined by
\be
g_{\mu\nu}=\bar{g}_{\mu\nu}+h_{\mu\nu}\,, 
\ee
the transformation laws are 
\ba 
\label{trans}
\Delta h_{ij}&=&2a^2H\ep_0\delta_{ij}-\bar g_{ik}\partial_j \ep^k-\bar g_{jk}\partial_i \ep^k \,,\\ \nn
&=&2a^2H\delta_{ij}\epsilon_0-2\ep_{(i,j)}\label{ij}\,,\\ 
\Delta h_{0i}&=&-\dot{\ep}_i-\partial_i \epsilon_0+2H\ep_i \,,\label{trans2}\\ 
\Delta h_{00}&=&-2\dot{\ep}_0\,,\label{trans3}
\ea
in which
\be
\ep_i\equiv a^2\ep^i\,.
\ee
The solid displacement fields obey
\be
\Delta \vpi^i=-\epsilon^i\,.
\ee
Finally, the energy-momentum tensor components transform as
\ba
\Delta \delta T_{00}&=&2\rho \dot{\ep}_0+\dot{\rho}\ep_0\hspace{3cm}\Rightarrow \Delta \delta \rho=\dot{\rho}\ep_0\,,\\
\Delta \delta T_{0i}&=&-p\dot{\ep}_i+\rho \partial_i \ep_0+2pH\ep_i\quad\quad\quad\Rightarrow \Delta \delta u_i=-\partial_i \ep_0\,,\\
\label{dTij}
\Delta \delta T_{ij}&=&-2p\ep_{(i,j)}+\partial_t(a^2\,p)\delta_{ij}\ep_0\,.
\ea
While the above transformation law for spacetime tensor only depend on $  \epsilon^{\mu} $, ambiguities arise when trying to define the transformation of scalar, vector and tensor perturbations because such a distinction, also known as Helmholz decomposition, is unique only at finite momentum. As an illustration, consider the transformation law \eqref{trans2} for
\be\label{finmon}
h_{0i}=a(t)\Big(\partial_i F+G_i\Big)\,.
\ee
One would like to derive the transformation of $  F $ and $  G_{i} $ separately by solving the differential equation
\be\label{inf}
\partial_{i}\partial_{i} \Delta F=\frac{1}{a}\partial_{i} \Delta h_{0i}
\ee
for $  \Delta F(x) $, and then using the solution into \eqref{finmon} to find $  \Delta G_{i} $. However, this equation admits a unique solution only for functions $ \Delta F(x) $ that vanish at spatial infinity. Conversely, allowing for functions that do not vanish at spatial infinity, there are infinitely many $  \Delta F(x) $ that satisfy \eqref{inf}, leading to infinitely many $ \Delta G_{i} $'s (see also \cite{Pajer:2017hmb} for more details). 


\subsection{Adiabatic modes in uniform-density gauge }

To study the properties of curvature perturbations on uniform-density slices, it is convenient to choose coordinates such that the constant time hypersurfaces coincide with the uniform-density slices, i.e.~  
\be
\label{UGG0}
\rho(t,\bfx)=\bar\rho(t)\,.
\ee
Spatial coordinates are chosen as follows
\be
\label{gaugecond}
g_{ij}=a(t)^2\exp(2\zeta)\Big(\exp(\gamma)\Big)_{ij}\,, \quad\quad \text{where}\quad\quad \gamma_{ii}=\partial_i \gamma_{ij}=0\,.
\ee
We will refer to this gauge choice as \textit{uniform-density} gauge. We borrow the ADM notation to refer to the metric perturbations in uniform-density gauge so as to make it distinct from an arbitrary gauge in the previous section, i.e.~we write
\be  
ds^2=-(1+N_1)^2dt^2+2(\partial_i \psi+N_i^V)dtdx^i+a(t)^2\exp(2\zeta)\Big(\exp(\gamma)\Big)_{ij}dx^idx^j\,.
\ee
In linear theory, \eqref{deltrho}, \eqref{deltaX} and \eqref{UGG0} imply the following relationship between $\zeta$ and $\vpi_L$ (the scalar phonon)
\be
\label{pzrel}
\nabla^2 \vpi_L=3\zeta\,.
\ee
Moreover, notice that in this gauge \eqref{deltrho}-\eqref{pijT} lead to
\ba
\delta u&=&-a^2\dot{\vpi}_L+\psi\,, \\ \nn
\delta u_i^V&=&-a^2\dot{\vpi}^V_i+N_i^V\,,\\ \nn
\pi^S&=&-4M_P^2\dot{H}c_T^2\,\vpi_L\,, \\ \nn
\pi^V_i&=&-2M_P^2\dot{H}c_T^2\,\vpi^V_i\,.
\ea

To derive adiabatic modes, we look for residual diffeomorphisms that are allowed by the uniform-density gauge choice. Assuming that $\bar\rho(t)$ is locally an invertible function of time, the uniform-density gauge excludes any temporal diff, $ \e^{0}=0$; spatial diffs, on the other hand, must obey the gauge condition \eqref{gaugecond}, i.e.~ 
\be
\label{resd}
\nabla^2 \ep_i=-\dfrac{1}{3}\partial_i \partial_k \ep_k\,.
\ee
Applying $\partial_i$ on both sides reveals $\nabla^2 \partial_i \ep_i=0$, therefore any non-trivial solution to this equation is large, i.e.~it cannot vanish at spatial infinity. According to \eqref{trans}-\eqref{dTij}, any spatial diff satisfying \eqref{resd} generates the following solution to the Einstein equations
\ba
\label{N1}
N_1&=&0\,,\\ 
\label{adz}
\zeta&=&-\dfrac{1}{3}\partial_k\ep^k\,,\\
\label{adn}
\partial_i \psi+N_i^V&=&-a^2\dot{\ep}^i\,,\\
\label{adp}
\partial_i \vpi_L+\vpi^V_i&=&-\ep^i\,.
\ea
In particular, $ \nabla^{2} \zeta\propto \nabla^{2}\partial_i \ep_i=0$. As an aside note that modes satisfying $ \nabla^{2}\zeta =0$ have been recently studied in \cite{Afshordi:2017use} under the name of ``cosmological zero modes''. It would be interesting to understand the connection between these results.


\subsubsection{All adiabatic modes}

Equation \eqref{resd} admits infinity many solutions, and each can be written as a finite order polynomials in $x^i$ (see e.g.~\cite{Hinterbichler:2013dpa}). However, not every generated perturbation is adiabatic---some do not continuously connect to any physical profile. To address the adiabaticity condition, it is useful to look at the following structure of the Einstein equations (${\cal E}_{\mu\nu}=0$) with respect to its $SO(3)$ decomposition:
\ba
&&(00): \quad {\cal E}_{00}=S^{(1)}=0\,,\\ \nn
&&(0i):\quad {\cal E}_i=\partial_iS^{(2)}+V_{i}^{(1)}=0\,,\qquad \partial_i V_{i}^{(1)}=0\,,\\ \nn
&&(ij):\quad {\cal E}_{ij}=S^{(3)}\delta_{ij}+\partial_i\partial_j\,S^{(4)}+2\partial_{(i}V_{j)}^{(2)}+T_{ij}=0\,,\qquad \partial_i V_{i}^{(2)}=\partial_i T_{ij}=T_{ii}=0\,.
\ea
At finite momentum, these equations straightforwardly imply $V^{(a)}=S^{(a)}=0$. However, at zero momentum, scalars, vectors and tensors can mimic each other. To ensure extension to finite momentum we need to demand
\be
\label{csteq}
S^{(2)}=0\quad (\text{or } V^{(1)}=0)\,,\text{ and}\quad S^{(3)}=V^{(2)}=0\,.
\ee
In the uniform-density gauge, these three equations are respectively
\ba
\label{0i}
&&\dot{H}\Big(-a^2\dot{\vpi}_L+\psi\Big)=HN_1-\dot{\zeta}\,,\\ 
\label{offij}
&&-4\dot{H}a^2c_T^2\,\vpi_L+N_1+\zeta+\dot{\psi}+H\psi=0\,, \\ 
\label{consv}
&&\dot{N}^V_i+HN_i^V-4\dot{H}a^2c_T^2\vpi^V_i=0\,,
\ea
and they constrain the time dependence of $\ep^i$. To see this, let us employ \eqref{N1} (which holds for any adiabatic solution) along with \eqref{0i} and \eqref{offij} so as to remove $\psi$ and arrive at the equation below
\be
\label{phiL}
\ddot{\vpi}_L+3H\dot{\vpi}_L-4\dot{H}c_T^2\vpi_L=\dfrac{1}{a^2\dot{H}}\ddot{\zeta}+\left(\dfrac{H}{a^2\dot{H}}-\dfrac{\ddot{H}}{a^2\dot{H}^2}\right)\dot{\zeta}-\dfrac{1}{a^2}\zeta\,.
\ee
Taking a Laplacian from both sides in conjunction with \eqref{pzrel} leads to a second order ODE for $\zeta$ 
\be
\ddot{\zeta}(t,\bfx)+3H\dot{\zeta}(t,\bfx)-4\dot{H}c_T^2\zeta(t,\bfx)=0\,, 
\ee
which is the same as in \eqref{cI}. As a result
\be
\label{zsol}
\zeta=c_1(\bfx)\cI_1(t)+c_2(\bfx)\cI_2(t)\,,
\ee
where $c_a(\bfx)$ are two arbitrary harmonic functions, i.e.~$\nabla^2 c_a=0$. 

Finally, we also need to implement \eqref{consv} which along with the adiabaticity conditions \eqref{adn}, \eqref{adp} and \eqref{offij} yields
\be
-\partial_i \zeta+a^2\Big(\ddot{\ep}^i+3H\dot{\ep}^i-4\dot{H}c_T^2\,\ep^i\Big)=0\,.
\ee
Plugging \eqref{zsol} inside this equation gives us 
\be
\ddot{\ep}^i+3H\dot{\ep}^i-4\dot{H}c_T^2\,\ep^i=\dfrac{1}{a(t)^2}\sum_a \partial_i c_a(\bfx)\cI_a(t)\,.
\ee
This equation together with \eqref{resd} implies the following form for the residual diffeomorphisms
\be
\label{finep}
\ep^i(t,\bfx)=\sum_a \ep^i_a(\bfx)\cI_a(t)+\sum_a \partial_i c_a(\bfx)\Big(\int dt'\,G(t,t')\dfrac{1}{a(t')^2}\cI_a(t')\Big)\,,
\ee
where $G(t,t')$ is the Green function of the equation \eqref{cI} and $\ep^a_i(\bx)$s are two arbitrary time-independent solutions to \eqref{resd}. Yet, there is another consistency relation between $c_a$ and $\ep^i_a$ enforced by \eqref{adz}, i.e.~
\be
c_a(\bfx)=-\dfrac{1}{3}\partial_i \ep^i_a(\bfx)\,.
\ee
In conclusion, by virtue of the equations \eqref{finep}, \eqref{phiL}, \eqref{0i}, \eqref{adn} and finally \eqref{adp}, we could systematically write down all possible adiabatic modes, as is summarized in Table \ref{ugadm}. 

\begin{table}
	\begin{center}
		\begin{tabular}{|c| c |}
			\hline
			&\\
			$\ep^i(t,\bfx)$&$\sum_a \ep^i_a(\bfx)\cI_a(t)-\dfrac{1}{3}\sum_a \partial_i \partial_j \ep^j_a(\bfx)\Big(\displaystyle\int dt'\,G(t,t')\dfrac{1}{a(t')^2}\cI_a(t')\Big)$\\
			&\\
			\hline
			&\\
			$\zeta$&$-\dfrac{1}{3}\sum\limits_a \partial_i \ep^i_a(\bfx)\cI_a(t)$\\
			&\\
			\hline
			&\\
			$\gamma_{ij}$&$-\sum\limits_a \Big(\partial_i \ep^j_a(\bfx)+\partial_j \ep^i_a(\bfx)-\dfrac{2}{3}\partial_k \ep^k_a(\bfx)\delta_{ij}\Big)\cI_a(t)$\\
			&$+\dfrac{2}{3}\sum\limits_a \partial_i \partial_j \partial_k\ep^k_a(\bfx)\Big(\displaystyle\int dt'\,G(t,t')\dfrac{1}{a(t')^2}\cI_a(t')\Big)$\\
			&\\
			\hline
			&\\
			$\vpi_L$& $\sum\limits_{a}\vpi_a(\bfx)\cI_a(t)-\dfrac{1}{3}\sum\limits_{a}\partial_i\ep^i_a(\bfx)\displaystyle\int dt'G(t,t') S_a(t')$\\
			&\\
			&$\nabla^2\vpi_a(\bfx)=-\partial_i \ep^i_a(\bfx)$\\
			&\\
			&$S_a(t')=\left[- \left( \dfrac{2H}{a(t')^2\dot{H}}+\dfrac{\ddot{H}}{a(t')^2\dot{H}^2} \right)\dot{\cI}_a(t')+\dfrac{1}{a(t')^2}(4c_T^2-1)\cI_a(t')\right]$\\
			&\\
			\hline
			&\\
			$\vpi^V_i$& $-\ep^i-\partial_i \vpi_L$\\
			&\\
			\hline
			 &\\
			$\psi$&$-\dfrac{1}{\dot{H}}\dot{\zeta}+a^2\dot{\vpi}_L$\\
			&\\
			\hline
			&\\
			$N_i^V$&$-a^2\dot{\ep}^i-\partial_i \psi$\\
			&\\
			\hline
		\end{tabular}
	\end{center}
\caption{\label{ugadm} All adiabatic modes for solid in the uniform density gauge. $\ep^i_a(\bfx)$ are two arbitrary solutions of \eqref{resd}, and ${\cal I}_a(t)$s are two linearly independent solutions of \eqref{cI}. $G(t,t')$ is the retarded Green function of \eqref{cI}} 
\end{table}
\subsubsection{Leading adiabatic modes}
The equation for residual diffs \eqref{resd} can be solved by Taylor expanding $\ep_a^i$ in spatial $\bfx$ and plugging back into Table \ref{ugadm}. In this section we present the leading adiabatic modes in the gradient expansion. Analogous to a perfect fluid coupled to gravity\cite{Pajer:2017hmb}, we will have two types of adiabatic modes: i) pure adiabatic modes, which are either of scalar, vector or tensor type, and ii) mixed adiabatic modes, which contains two types of SVT at the same time. By definition, a mixed adiabatic mode cannot be made pure by applying any additional large diff. 
\subsubsection*{$\diamond$ Pure scalar: curvature mode}
A non-trivial solutions to \eqref{resd} is a simple spatial scaling, i.e.~
\be
\ep^i_a(\bfx)=\lambda_a x^i\,,
\ee
leading to a spatially homogeneous curvature perturbation
\be
\label{purecurve}
\zeta=-\sum_a \lambda_a \cI_a(t)\,.
\ee
If, in addition, we choose 
\be
\vpi_a(\bfx)=-\dfrac{1}{2}\lambda_a \bfx^2\,,
\ee
to fulfill $\nabla^2 \vpi_a=-\partial_i \ep^i_a$, we discover
\ba
\vpi_L&=&-\sum_a \dfrac{1}{2}\lambda_a \bfx^2 \cI_a(t)-\sum_a \lambda_a \int dt'G(t,t')S_a(t')\,,\\ \nn
\psi&=&\sum_a \lambda_a \Big(\dfrac{1}{\dot{H}}-\dfrac{1}{2}a^2\bfx^2\Big)\dot{\cI}(t)\,.
\ea
Above, we have defined
\be
\label{Sa}
S_a(t')=\left(-(\dfrac{2H}{a(t')^2\dot{H}}+\dfrac{\ddot{H}}{a(t')^2\dot{H}^2})\dot{\cI}_a(t')+\dfrac{1}{a(t')^2}(4c_T^2-1)\cI_a(t')\right)\,.
\ee
Notice that vector and tensor perturbations vanish for this particular mode.

Naively, due to the nonzero anisotropic stress of the solid and in contrast with standard inflation, one might expect that any long mode have a physical effect on local physics. It might therefore seem counter-intuitive that a long wavelength longitudinal phonon is adiabatic, i.e.~locally indistinguishable from a change of coordinates.  However, the $\bfx^2$ structure that appears in $\vpi_L$ and $\psi$ makes it impossible to promote this adiabatic mode to a plane wave. Rather, this adiabatic mode arises as the radially homogeneous limit of a generic spherical perturbation. Indeed, it has been observed in \cite{Bordin:2017ozj} that such a spherical long mode in a solid is locally unobservable, in agreement with our findings.

 
\subsection*{$\diamond$ Pure scalar: isocurvature mode}
As a trivial possibility, let us impose $\ep^i_a=0$ and also take
\be
\phi_a(\bfx)=c_a\,.
\ee
These choices induce a scalar mode
\ba
\zeta&=&0\,,\\
\vpi_L&=&\sum_a c_a\cI_a(t)\,,\\ 
\psi&=&a^2\dot{H}\sum_a c_a\dot{\cI}_a(t)\,.
\ea
Quite remarkably, this adiabatic mode is not associated with any residual diffeomorphism, i.e.~$\ep^i(t,\bfx)=0$. This is consistent with the Einstein equations, since an $\bfx$ independent $\vpi_L$ and $\psi$ do not generate any perturbation in the metric nor in the energy-momentum tensor. Yet, the extendibility to finite momentum determines the non-trivial time dependence of this mode. 


\subsection*{$\diamond$ Pure scalar: isocurvature gradient mode}
Inserting an $\bfx$ independent diff, i.e.~
\be
\ep^i=\sum_a c^i_a\cI_a(t)\,,
\ee
along with 
\be
\vpi_a(\bfx)=-c^i_ax^i\,,
\ee
generates a gradient mode in scalars, given by
\ba
\zeta&=&0\,,\\
\vpi_L&=&\sum_a c^i_ax^i\,\cI_a(t)\,,\\
\psi&=&a^2\sum_a c^i_ax^i\,\dot{\cI}_a(t)\,.
\ea
Notice that this and the previous isocurvature adiabatic mode are not associated with any soft theorems, as they do not induce any transformation on the observable quantities such as $\zeta$ or $\partial_i \phi^j$.


\subsection*{$\diamond$ Pure vector mode}

By means of a time dependent translation, i.e.~by choosing 
\be\label{vector}
\ep^i_a(\bfx)=d^i_a\,,
\ee
a pure vector perturbation can be generated  
\ba
\vpi^V_i&=&\sum_{a=1,2}d_a^i \cI_a(t)\,,\\
N^V_i&=&-a^2\sum_{a=1,2}d_a^i \dot{\cI}_a(t)\,.
\ea


 \subsection*{$\diamond$ Mixed adiabatic mode: ${\cal O}(\bfx^0)$ in tensor, ${\cal O}(\bfx^2)$ in scalars (or ${\cal O}(\bfx)$ in vectors)}
Assuming 
 \be
\ep^i_a=-\tilde{\omega}_{ij}x^j\,,~~~~~\tilde{\omega}_{ij}=\tilde{\omega}_{ji}\,~~~\text{and}~~~ \tilde{\omega}_{kk}=0\,,
\ee
together with 
\be
\vpi_a(\bfx)=\dfrac{1}{2}\tilde{\omega}_{ij}x^ix^j\,,
\ee
generates the following mixed mode
\ba
\label{tsmixed}
\gamma_{ij}&=&2\sum_{a=1,2}\tilde{\omega}^a_{ij}\cI_a(t)\,,\\ \nn
\vpi_L&=&\dfrac{1}{2}\sum_{a=1,2}\tilde{\omega}^a_{ij} {\cal I}_a(t)\, x^i x^j\,,\\ \nn
\psi&=&\dfrac{1}{2}a^2\sum_{a=1,2}\tilde{\omega}^a_{ij} \dot{\cI}_a(t)\, x^i x^j\,.
\ea
By replacing the above $\phi_a(\bfx)$ with zero, it is possible to trade the scalar mode for a gradient in vectors, given by 
\ba
\vpi_V^i&=&\sum_{a=1,2}\tilde{\omega}^a_{ij}x^j \cI_a(t)\,,\\
N_i^V&=&a^2 \sum_{a=1,2}\tilde{\omega}^a_{ij}x^j \dot{\cI}_a(t)\,.
\ea


\subsection*{$\diamond$ Mixed adiabatic mode: ${\cal O}(\bfx,\bfx^3)$ in scalars, ${\cal O}(\bfx^0,\bfx^2)$ in vectors}

Here we derive a gradient adiabatic mode for $\zeta$, in order to illuminate how the machinery extends beyond the leading order in gradient expansion.  
Inserting special conformal transformation for $\ep^i_a(\bfx)$, i.e.
\be
\ep^i_a=-x^ix^jb^j_a+\dfrac{1}{2}b^i_a x^jx^j\,,
\ea
we arrive at the following diffeomorphism
\be
\ep^i=\sum_a \Big(-x^ix^jb^j_a+\dfrac{1}{2}b^i_a x^jx^j\Big)\cI_a(t)+\sum_a b_a^i\Big(\displaystyle\int dt'\,G(t,t')\dfrac{1}{a(t')^2}\cI_a(t')\Big)\,.
\ee
In turn, this leads to a mixed adiabatic mode as below
\ba
\zeta&=&\sum_a b_a^ix^i \cI(t)\,,\\
\vpi_L&=&\dfrac{3}{10}\sum_a b^i_ax^i \bfx^2\cI_a(t)+\sum_a b^i_ax^i\int dt'G(t,t')S_a(t')\,,\\ 
\vpi_V^i&=&\dfrac{2}{5}\sum_a \Big(b^j_ax^j x^i-2b^i_a \bfx^2\Big)\cI_a(t)-\sum_a b^i_a \int dt'G(t,t')\Big(\dfrac{\cI_1(t')}{a(t')^2}+S_a(t')\Big)\,.
\ea

Since vectors decay at late times, it might seem reasonable to trade vectors for tensors, as we did for the previous adiabatic mode. However, according to Table \ref{ugadm}, Ad.Mo. that do not contain vectors have spatially constant $\zeta$, hence making this task unreachable.
Another possibility is  to generate a gradient mode in $\partial_i \vpi^j$ in place of $\zeta$, and this will lead to the next Ad. Mo.
\subsection*{$\diamond$ Mixed adiabatic mode: ${\cal O}(\bfx^3)$ in scalars, ${\cal O}(\bfx)$ in tensors}
Let us choose 
\be
\vpi_a(\bfx)=\dfrac{1}{3!}M^a_{ijk}x^ix^jx^k\,,~~~~ M^a_{ijk}=\text{totally symmetric and traceless}\,,
\ee
which ensures that $\zeta=0$. In order to avoid the appearance of vectors we are enforced to take
\be
\ep^i_a(\bfx)=-\dfrac{1}{2}\sum_a M^a_{ijk}x^jx^k\,.
\ee
The generated adiabatic mode will be given by
\ba
\label{gradient}
\partial_i\partial_j\phi_L&=&\sum_a M^a_{ijk}x^k \cI_a(t)\,,\\ \nn
\gamma_{ij}&=&\sum_a \cI_a(t)\,M^a_{ijk}x^k\,,\\ \nn
\psi&=&a^2\sum_a \dfrac{1}{3!}M^a_{ijk}x^ix^jx^k \dot{\cI}_a(t)\,.
\ea


\subsection{Adiabatic modes in uniform-density unitary gauge}

From our introductory discussions on adiabatic modes and soft theorems, it might seem that residual diffs play a central role in their derivation. Nevertheless, it is always possible to choose a local gauge that does not possess any residual diff whatsoever. Yet, even in those gauges adiabatic modes exist and  represent solutions that become ``nothing" in the adiabatic limit. The set up of a solid admits a gauge of this sort, which we will call the uniform-density unitary gauge (UDU) and is defined by 
\be
\vpi^i=0\,~~\text{and}~~ \rho=\rho(t)\,, \quad \text{(UDU gauge)}\,.
\ee
For the metric and the energy-momentum tensor we use the conventions as in \eqref{mconv} and \eqref{thermo}. According to \eqref{deltrho} and \eqref{deltaX}, the homogeneity of the time slicing in this gauge enforces that
\be
\label{ABrel}
A=-\dfrac{1}{3}\nabla^2B\,.
\ee

No (large) coordinate change is allowed after fixing the foliation of spacetime to uniform-density unitary gauge. Consequently, adiabatic modes must obey
\be
\label{vanish}
h_{\mu\nu}=0\,, 
\ee
or in detail 
\ba
\label{g00}
&&E=0\,,\\ 
\label{g0i}
&&G_i+\partial_i F=0\,,\\
\label{gij}
&&(\partial_i \partial_j-\dfrac{1}{3}\delta_{ij}\nabla^2)B+\partial_{(i}C^V_{j)}+\gamma_{ij}=0\,.
\ea
Note that in this gauge all degrees of freedom are eaten by $h_{\mu\nu}$ , thus $\delta T_{\mu\nu}=0$ does not provide any new information. 
If we were to solely deal with physical quantities for which the SVT decomposition is uniquely determined by $h_{\mu\nu}$, we would have simply arrived at 
\be
A=B=C_i=\gamma_{ij}=\delta \rho=....=0\,. 
\ee
However, with large perturbations it is always possible to generate non-zero scalar, vector or tensor solutions that add up to zero. As a toy example let us take 
\be
A=1\,,~~ B=-\dfrac{1}{2}\bfx^2\,,~~C_i=\gamma_{ij}=0\,.
\ee
For this non-trivial ansatz it is easy to check that 
\be
h_{ij}=a^2A\delta_{ij}+a^2\partial_i\partial_j B=0\,.
\ee

In conclusion, the only restriction on adiabatic modes in  uniform- density unitary is imposed by \eqref{vanish} and the adiabaticity condition \eqref{csteq}, which becomes
\ba
\label{dynB}
&&\ddot{B}+3H\dot{B}+\dfrac{1}{3a^2}\nabla^2 B-4\dot{H}c_T^2\,B-\dfrac{2}{a}\dot{F}-\dfrac{4H}{a}F=0\,,\\ 
\label{0iscalar}
&&\dot{H}\,a\,F=\dfrac{1}{2}H\,E+\dfrac{1}{6}\nabla^2\dot{B}\,,\\
\label{Cit}
&&\ddot{C}_i+3H\dot{C}_i-4\dot{H}c_T^2\,C_i-\dfrac{1}{a}\dot{G}_i-\dfrac{2H}{a}G_i=0\,,\\ 
\label{vecij}
&&4\dot{H}\,a\,G_i^V=-\dfrac{1}{a}\nabla^2 G_i+\nabla^2 \dot{C}_i\,.
\ea


\subsubsection{All adiabatic modes}
Applying the operator $\partial_i\partial_j$ on the both sides of \eqref{gij} reveals that 
\be
\label{nab4}
\nabla^4 B=\nabla^2 A=0\,. 
\ee
Let us act with the Laplacian operator $\nabla^2$ on \eqref{dynB}. Along with \eqref{nab4},  \eqref{0iscalar} and \eqref{ABrel} this will yield the following equation for $A(t,\bfx)$ 
\be
\label{Aode}
\ddot{A}+3H\dot{A}-4\dot{H}c_T^2\,A=0\,,
\ee
which is the same equation as \eqref{cI}. 

Taking the derivative $\partial_i$ of \eqref{gij} and \eqref{g0i} results in 
\be
\label{consAC}
\nabla^2C_i=2\partial_i A\,,~~~\nabla^2 G_i=\nabla^2F=0\,,
\ee
From which, using also \eqref{vecij}, we infer that
\be
\label{Gi}
G_i=\dfrac{1}{2\dot{H}a}\partial_i \dot{A}\,.
\ee

Now we are ready to construct every adiabatic mode in this gauge. First of all, equations \eqref{nab4} and \eqref{Aode} imply that 
\be
A(t,\bfx)=\sum_a f_a(\bfx)\cI_a(t)\,,~~~~\nabla^2 f_a(\bfx)=0\,,
\ee
Hence from \eqref{Gi} we find
\be
G_i=\dfrac{1}{2\dot{H}a}\sum_a \partial_i f_a(\bfx)\dot{\cI}_a(t)\,.
\ee
The equation \eqref{Cit} accompanied by \eqref{consAC} and \eqref{cI} reveals that
\be
C_i=\sum_a C_i^a(\bfx)\cI_a(t)+\dfrac{1}{2}\sum_a \partial_i f_a(\bfx) \int^t dt'\,G(t,t')\left(S_a(t')+\dfrac{\cI_a(t')}{a(t')^2}\right)\,,
\ee
where $S_a(t), a=1,2$ are two functions defined in \eqref{Sa} and we have
\be
\nabla^2 C^a_i(\bfx)=2\partial_i f_a(\bfx)\,\quad\text{and}\quad \partial_i C^a_i=0\,.
\ee
Finally, employing \eqref{dynB} and \eqref{gij} respectively shows us that
\ba
B&=&\sum_a B_a(\bfx)\cI_a(t)-\sum_a f_a(\bfx)\int^t dt' G(t,t')S_a(t')\,,\\ \nn
\gamma_{ij}&=&-2 \sum_a \partial_{(i} C^a_{j)}(\bfx)\cI_a(t)-\delta_{ij}\sum_a f_a(\bfx)\cI_a(t)\\ \nn
&&-\sum_a \partial_i \partial_j f_a(\bfx)\int dt'\, G(t,t')\dfrac{\cI_a(t')}{a(t')^2}-\sum_a \partial_i \partial_j B_a(\bfx)\cI_a(t)\,.
\ea
where
\be
\nabla^2 B_a(\bfx)=-3f_a(\bfx)\,.
\ee


\subsubsection{Leading adiabatic modes}

For the sake of brevity, in this section we only report the two most important adiabatic modes that lead to soft theorems, namely a pure scalar and a mixed tensor-scalar (tensor-vector) mode.


\subsection*{$\diamond$ Pure scalar: curvature mode}
Consider the following choice of parameters
\ba
f_a(\bfx)=c_a\,,\quad\quad
B_a(\bfx)=-\dfrac{1}{2}c_a \bfx^2\,,\quad\quad
C^a_i(\bfx)=0\,.
\ea
This will generate
\ba
A&=&\sum_a c_a \,\cI_a(t)\,,\\ \nn
B&=&-\dfrac{1}{2}\bfx^2\sum_a c_a \cI_a(t)+\bar{B}(t)\,,\\ \nn
F&=&-\dfrac{1}{2}\sum_a c_a \dfrac{\dot{\cI}(t)}{a\,H}\,,
\ea
in which 
\ba
\bar{B}(t)&=&-\sum_a c_a(\bfx)\int^t dt' G(t,t')S_a(t')\,.
\ea
Recall that in uniform-density unitary gauge
\be
\zeta=\dfrac{A}{2}\,.
\ee
A simple computation shows that this adiabatic mode is identical to \eqref{purecurve}. 
 \subsection*{$\diamond$ Mixed adiabatic mode: ${\cal O}(\bfx^0)$ in tensor, ${\cal O}(\bfx^2)$ in scalars (or ${\cal O}(\bfx)$ in vectors)}
Let us input
\ba
f_a(\bfx)=0\,,\quad\quad
B_a(\bfx)=0\,,\quad\quad
C^a_i(\bfx)=\tilde{\omega}^a_{ij}x^j\,.
\ea
generating a mixture of tensors and vectors, i.e.~
\ba
\gamma_{ij}&=&-2\sum_a \tilde{\omega}^a_{ij}\,\cI_a(t)\,,\\ \nn
C_i&=&\sum_a \,\tilde{\omega}^a_{ij}\,x^j\,\cI_a(t)\,.
\ea
We can do better by inserting 
\ba
&&f_a(\bfx)=0\,,\quad\quad
B_a(\bfx)=\tilde{\omega}^a_{ij}x^ix^j\,,\quad\quad \tilde{\omega}_{ii}=0\,,\\ \nn
&&\qquad\qquad\qquad C^a_i=0\,,
\ea
so as to get
\ba
A&=&0\,,\\ \nn
B&=&\sum_a \tilde{\omega}^a_{ij}x^ix^j\,\cI_a(t)\,,\\ \nn
\gamma_{ij}&=&-2\sum_a \tilde{\omega}^a_{ij}\,\cI_a(t)\,.
\ea


\section{Solid soft theorems}\label{sec3}

In this section, we derive solid soft theorems, namely relations among $n$- and $n+1$-point functions, for which the momentum of one of the $n+1$ fields goes to zero. We do this by combining the asymptotic symmetries of the previous section with the Operator Product Expansion (OPE) \cite{Weinberg:1996kr}. Among all soft bispectra, we focus on evaluating $\langle \gamma \zeta\zeta\rangle$, $\la \zeta\zeta\zeta\ra$ and $\la \vpi_V\zeta\zeta\ra$. Of course, the first two are the most promising ones from an observational perspective. Finally, we illuminate how the logic extends beyond leading order in momentum expansion---as an illustration, we derive a soft theorem for trispectrum implied by the gradient adiabatic mode in \eqref{gradient}.

Let us outline our strategy, which proceeds along the lines of \cite{Finelli:2017fml,Assassi:2012et,Kehagias:2012pd}. The first observation is that every adiabatic mode is related to a nonlinearly realized asymptotic symmetry that generates a solution that contains both a linear and a nonlinear term (the latter being the adiabatic mode solution). Let us consider the pure scalar adiabatic mode as an example. The corresponding asymptotic symmetry is the dilation symmetry $x^{i} \to x^{i}+\epsilon^{i}$, with $\epsilon^{i}=-\sum_{a}c_{a}I_{a}(t)\x^{i}$. The corresponding charge acts on scalar fields by generating the adiabatic mode as well as a linear transformation. For instance, when the dilation charge acts on $\zeta$, we find
\begin{align}\label{Qonzeta}
\left[Q_{dil},\zeta(x)\right]=\frac{1}{3}\partial_{i}\epsilon^{i}+\epsilon^{i}\partial_{i}\zeta(x).
\end{align}

Then we assume the OPE to hold, such that we can write the product of two nearby fields as an expansion in all locally observable operators allowed by the symmetries of the theory. For instance, the expansion of two nearby $\zeta$ operators in Fourier space will give something of the form
\begin{align}\label{charge0}
\zeta_{\bf{k}-\frac{1}{2}\q}\zeta_{-\bf{k}-\frac{1}{2}\q}\xrightarrow{\q\to 0}P(k)(2\pi)^{3}\delta^{3}(\q)+f(k)\zeta_{-\q}+\cdots \, ,
\end{align} 
for some function $f(k)$, and dots refer to any locally observable operator allowed by the symmetries. The trick is then to act with $Q_{dil}$ on both sides of the OPE and take the expectation value. Due to the linear-nonlinear structure of \eqref{charge0}, this leads to a nontrivial relation, which allows us to fix some or all of the coefficients of the OPE. 

Having fixed the coefficients of the OPE, we can finally insert the OPE in for instance the squeezed bispectrum to obtain soft theorems. Below, we explicitly go through these steps for all solid adiabatic modes. We highlight which coefficients can be fixed in each case, and when spherical averaging is required. 
 

\subsection{Soft scalar theorem}

In the previous section, we showed that the solid supports two pure scalar adiabatic modes. Both of them can be defined in terms of the solution for $\varphi_{L}$, from which for instance $\zeta$ follows directly. The local observables are then simply derivatives of $\varphi_{L}$. See Appendix \ref{app:derivatives} for the rationale behind the number of derivatives per field we consider. Up to second order in derivatives, this yields for the OPE
\begin{align}\label{OPE}
\lim_{q\to 0}\zeta_{\bfk-\bfq/2}\zeta_{-\bfk-\bfq/2}=& \, P(k)(2\pi)^{3}\delta^{3}(q)
+f^{ij}(\bfk)(\partial_i \partial_j \vpi_L)_{-\bfq}+g^{ij}(\bfk)(\partial_i \partial_j \dot{\vpi}_L)_{-\bfq} \nn \\ &+\mathcal{O}(q^2 \zeta, \zeta^2,\gamma,\vpi^V_i,\vpi_L^2)\,,
\end{align}
where corrections $  \O(q\zeta) $ are forbidden by the parity argument given around \eqref{parity}. The expectation value of the action of the dilation charge on the operator on the right is
\begin{align}
\la[Q_{dil},\left(\partial_i \partial_j \vpi_L\right)_\bfq]\ra=\sum_{a=1,2}c_a\cI_a(t)\delta_{ij}(2\pi)^3 \delta^3(\bfq)\,.
\end{align}  

Since the charge is time-independent, the action of the charge on time derivative of the above is straightforwardly obtained as the time derivative of the commutator. Applying the same commutator and expectation value on both sides of the OPE, we get 
\begin{align}\label{lhs}
\langle\left[Q_{dil},\zeta_{\bfk-\frac{\bfq}{2}}\zeta_{-\bfk-\frac{\bfq}{2}}\right]\rangle &= \langle\left[Q_{dil},\zeta_{\bfk-\frac{\bfq}{2}}\right]\zeta_{-\bfk-\frac{\bfq}{2}}\rangle+\langle \zeta_{\bfk-\frac{\bfq}{2}}\left[Q_{dil},\zeta_{-\bfk-\frac{\bfq}{2}}\right]\rangle \nonumber\\
&=\sum_{a=1,2}c_a\cI_a(t)\delta^{3}\left(\bfq\right)\left(3+\bfk \cdot \partial_{\bfk}\right)P(k).
\end{align}
Matching the time dependence and soft momentum dependence, one finds  
\be\label{fijk}
 f^{ij}(k)\delta_{ij}=(n_{s}-1)P(k)\,,\quad g^{ij}(k)\delta_{ij}=0\,.
\ee
This does not fully specify $f^{ij}$ or $g^{ij}$. In the scalar sector, statistical isotropy allows us to decompose
\ba
\label{sdecomp}
f^{ij}(k)=f_{\text{aniso}}\left(\hat{k}^{i}\hat{k}^{j}-\frac{1}{3}\delta_{ij}\right)+f_{\text{iso}}\delta_{ij}\,,\\ \nn
g^{ij}(k)=g_{\text{aniso}}\left(\hat{k}^{i}\hat{k}^{j}-\frac{1}{3}\delta_{ij}\right)+g_{\text{iso}}\delta_{ij}\,.
\ea
From \eqref{fijk}, we thus find
\be
f_{\text{iso}}=\frac{1}{3}(n_{s}-1)P(k) \,, \quad g_{\text{iso}}=0\,,
\ee
but we are \textit{not} able to constrain $f_{\text{aniso}}$ or $g_\text{aniso}$. The squeezed bispectrum then reads
\ba
&&\lim_{\bfq\to 0}\langle \zeta_{\bfq} \zeta_{\bfk+\bfq/2}\zeta_{-\bfk+\bfq/2}\rangle^{\prime} =\left[(n_{s}-1)+3f_{\text{aniso}}\left(\cos^{2}\theta-\frac{1}{3}\right)\right]P(k)P(q)\\ \nn
&&\hspace{6cm}+\dfrac{3}{2}g_{\text{aniso}}\left(\cos^{2}\theta-\frac{1}{3}\right)P(k)\dot{P}(q)\,.
\ea
In terms of the expansion \eqref{Legendreexp}, the soft theorem can be stated as
\be
a_{0}^{(0)}=(n_s-1)\,.
\ee
We have therefore re-derived the result of \cite{Bordin:2017ozj}: only upon angular averaging the squeezed bispectrum (which kills the $f_{\text{aniso}}$ and $  g_{aniso} $ terms) do we reach the well known single field soft theorem \cite{Maldacena:2002vr}, 
\be
\label{Malda}
\dfrac{1}{2}\int_{-1}^{+1} d\cos\theta \dfrac{1}{P_\zeta(q)}\la \zeta_\bfq \zeta_{\bfk-\frac{1}{2}\bfq}\zeta_{-\bfk-\frac{1}{2}\bfq}\ra=(n_s-1)P_\zeta(k)\,.
\ee
In fact, by using the same OPE method, it is easy to see that this soft theorem generalizes to soft (n+1)-spectra, i.e.~
\be
\boxed{
\lim_{q\to 0}\int \dfrac{d^2\hat{q}}{4\pi}\,\dfrac{1}{P_\zeta(q)}\la \,\zeta_{-\bfq}\prod_{a=1}^{n}\zeta_{\bfk_a+\bfq/n}\ra'=\left[ 3(n-1)+\sum_a\,\bfk_a.\frac{\partial}{\partial \bfk_a} \right] \,\la \,\prod_{a=1}^{n}\zeta_{\bfk_a+\bfq/n}\ra'\,.}
\ee


\subsection{Soft scalar theorem in uniform-density unitary  gauge}

As we discussed, in uniform-density unitary gauge there are no residual large diffeomorphisms. Yet, there are adiabatic modes and the scalar soft theorem discussed in the previous subsection can also be derived in this gauge as well. In this subsection we show in some details how this works. The key observation is that the symmetry creates both a linear and a nonlinear term when it acts on the fields in uniform-density gauge, whereas the linear part is absent in uniform-density unitary gauge. 

This means that when we let the symmetry act on the left hand side of the OPE, and take the expectation value, it vanishes. As a consequence, the right hand side has to cancel among itself, which only allows for a quadrupolar contribution to the squeezed limit, which vanishes upon angular averaging. 

In more detail, let
\begin{align}
A_{\bfk+\frac{\bfq}{2}}A_{-\bfk+\frac{\bfq}{2}}=&(2\pi)^3\delta^3(\bfq)\la A_\bfk A_{-\bfk}\ra'+
+F^{ij}(\bfk)(\partial_i \partial_j B)_\bfq+G^{ij}(\bfk)(\partial_i \partial_j \dot{B})_\bfq+...\, ,
\end{align}
where we expanded in terms of $B$ for convenience, since it is directly related to $A$ by \eqref{ABrel}. The derivation the soft theorem is then analogous to the previous subsection, with the exception that $\la[Q_S,A_{\bfk+\frac{\bfq}{2}}A_{-\bfk+\frac{\bfq}{2}}]\ra=0$. If we write 
\ba
F^{ij}(k)&=&F_{\text{aniso}}\left(\hat{k}^{i}\hat{k}^{j}-\frac{1}{3}\delta_{ij}\right)+F_{\text{iso}}\delta_{ij}\, ,\\ \nn
G^{ij}(k)&=&G_{\text{aniso}}\left(\hat{k}^{i}\hat{k}^{j}-\frac{1}{3}\delta_{ij}\right)+G_{\text{iso}}\delta_{ij}
\ea
and use that
\begin{align}
[Q_S,(\partial_i \partial_j B)_\bfq]=-\sum_{a=1,2}c_a\cI_a(t)\delta_{ij}(2\pi)^3 \delta^3(\bfq),
\end{align}
we immediately conclude that
\ba
\lim_{\bfq\to 0}\langle A_{\bfq} A_{\bfk+\bfq/2}A_{-\bfk+\bfq/2}\rangle^{\prime} &=& 3F_{\text{aniso}}\left(\cos^{2}\theta-\frac{1}{3}\right)P_A(q)P_A(k) \, \\ \nn
&&+\dfrac{3}{2} G_{\text{iso}}\left(\cos^{2}\theta-\frac{1}{3}\right)\dot{P}_A(q)P_A(k)\,,
\ea
which vanishes upon angular averaging
\ba
\int d(\cos\theta)\,\lim_{\bfq\to 0}\langle A_{\bfq} A_{\bfk+\bfq/2}A_{-\bfk+\bfq/2}\rangle^{\prime}=0\,. \label{zero}
\ea
Perhaps contrary to its appearance, this perfectly agrees with the result for the $\zeta$ three-point function. In fact, as we show in Appendix \ref{app:transformation}, accounting for the second order relation between $A$ and $\zeta$ the two soft theorems are equivalent. This proves that the final soft theorem for $\zeta$ is gauge invariant as it should be.


\subsection{Mixed vector-tensor soft theorems}

Mixed adiabatic modes lead to mixed soft theorems which combine soft modes of different SVT character. We found a mixed vector-tensor and a scalar-tensor mode. In this and the next subsection, we derive the corresponding soft theorems. As we will see, unlike for soft scalars, these do not involve any angular averaging. 

In the absence of long scalar perturbations, the OPE can be written as
\begin{align}
\nn
\zeta_{\bfk+\bfq/2}\zeta_{-\bfk+\bfq/2}=&f^{ij}_T(\bfk)\gamma_{ij}(\bfq)+g^{ij}_T(\bfk)\dot{\gamma}_{ij}(\bfq)\\ 
\label{OPE}
&+f^{ij}_V (\bfk)(\partial_i \vpi_j^V)_\bfq+g^{ij}_V(\bfk) (\partial_i \dot{\vpi}_j^V)_\bfq+{\cal O}(\gamma^2, \zeta^2, \vpi^2,q^2\gamma,q^{2}\zeta)
\end{align}
Again the number of spatial derivatives acting on $\vpi$ is given by the shift symmetry of the theory, see Appendix \ref{app:derivatives}, and again the next order derivative corrections appear only at order $  \O(q^{2}\zeta,q^{2}\gamma) $ by the parity argument around \eqref{parity}. As before, we care about the linear part of the symmetry generator for the left hand side, whereas we care about the nonlinear part on the right hand side. For the gauge transformation corresponding to the mixed modes, i.e.
\begin{align}
\ep^0=0\,,\quad \ep^i=\sum_{a}\tilde{\omega}_{ij}^a \cI_a(t)x^j\, , 
\end{align} 
the relevant commutators are
\begin{align}
\label{mixedg}
[Q,\zeta_{\bfk}]_{\text{lin}}=&-\sum_{a}\tilde{\omega}_{ij}^a \cI_a(t)k^i\dfrac{\partial}{\partial k^j}\zeta_{\bfk} \,,\\ \nn 
[Q,\gamma_{ij}(\bfq)]_{\text{nlin}}=&2\sum_{a=1,2}\tilde{\omega}^a_{ij}\cI_a(t)(2\pi)^3\delta^3(\bfq)\,,\\ \nn
[Q,(\partial_i \vpi_j^V)_{\bfq}]_{\text{nlin}}=&\sum_{a=1,2}\tilde{\omega}^a_{ij}\cI_a(t)(2\pi)^3\delta^3(\bfq)\,.
\end{align}
Using that 
\begin{align}\label{rhsmixed}
\left (k_1^i\dfrac{\partial}{\partial k_1^j}+k_2^i\dfrac{\partial}{\partial k_2^j}\right)\langle\zeta_{\bfk_1}\zeta_{\bfk_2}\rangle= \dfrac{k^ik^j}{k}\dfrac{\partial P(k)}{\partial k}(2\pi)^3\delta^3(\bfq)-P(k)\delta_{ij}(2\pi)^3\delta^3(\bfq) \, ,
\end{align}
we find 
\begin{align}\label{lhsmixed}
\la[Q,\zeta_{k_1}\zeta_{k_2}]\ra= -\sum_{a}\tilde{\omega}_{ij}^a \cI_a(t)\left (\dfrac{k^ik^j}{k}\dfrac{\partial P(k)}{\partial k}(2\pi)^3\delta^3(\bfq)\right) \, ,
\end{align}
where we used that $\tilde{\omega}_{ij}^a$ is traceless. After acting with the charge and taking the expectation value on the right hand side, it is useful to decompose the terms on the right in terms of their spatial structure as before. Since the vectors and tensors are traceless, we can immediately write 
\begin{align}\label{mixeddecomp}
f^{ij}_\alpha(\bfk)=&f_\alpha(k)\left(\hat{k}^i\hat{k}^j-\dfrac{1}{3}\delta_{ij}\right)\,,\\ 
g^{ij}_\alpha(\bfk)=&g_\alpha(k)\left(\hat{k}^i\hat{k}^j-\dfrac{1}{3}\delta_{ij}\right)\,,
\end{align}
where $\alpha=T,V$. Matching the time-dependence then implies 
\begin{align}
\label{consist}
2g_T(k)+g_V(k)=&0\,,\\ \nn
2f_T(k)+f_V(k)=&-k\partial_{k}P_\zeta(k)\,.
\end{align}
In summary, the final OPE reads
\ba
\zeta_{\bfk+\bfq/2}\zeta_{-\bfk+\bfq/2}&=&f_T(k)\hat{k}^i\hat{k}^j\,\gamma_{ij}(\bfq)+i\,f_V(k)\hat{k}^i\hat{k}^j q_i\vpi^V_j(\bfq)+\\ 
\nn
&&+g_T(k) \hat{k}^i\hat{k}^j \dot{\gamma}_{ij}(\bfq)+ig_V(k)\hat{k}^i\hat{k}^j\,q_i\,\dot{\vpi}^V_j(\bfq)\, ,
\ea
subject to \eqref{consist}. The mixed soft theorem is then obtained by correlating these expressions with long vector and tensor fields,
\begin{align}
\label{gssvss}
\la \gamma^s(-\bfq)\zeta_{\bfk+\bfq/2}\zeta_{-\bfk+\bfq/2}\ra'=& \, \hat{k}^i\hat{k}^j\ep^s_{ij}(\hat{q})\,\left( f_T(k)\,P_{\gamma}(t,q)+\dfrac{1}{2}g_T(k) \dot{P}_{\gamma}(t,q)\right)\,, \nn \\
\la \vpi^s(-\bfq)\zeta_{\bfk+\bfq/2}\zeta_{-\bfk+\bfq/2}\ra'= & \, iq\,\hat{k}^i\ep^s_i(\hat{q})\,\cos\theta\,\left(f_V(k)P_V(t,q)+\dfrac{1}{2}g_V(k)\dot{P}_V(t,q)\right)\, ,
\end{align}
In order to infer any soft theorem by inserting \eqref{consist} in \eqref{gssvss}, we need to have
\be
\lim\limits_{q\to 0} \left(\dfrac{\dot{P}_{\gamma}(t,q)}{P_{\gamma}(t,q)}-\dfrac{\dot{P}_V(t,q)}{P_V(t,q)}\right)=0\,.
\ee
For solid, this is satisfied provided\footnote{This also means that the superhorizon modes of the solid are asymptotically classical.} 
\be
\lim\limits_{t\to \infty}\dfrac{I_2(t)}{I_1(t)}=0\,,
\ee
i.e.~if one of the two solutions of \eqref{cI} decays with time with respect to the other as it often happens for cosmological perturbations. If this is the case, the mixed soft theorem turns out to be 
\ba\label{st1}
\boxed{
2b^{(0)}_{0}(k)+\,c_{1}^{(1)}(k)=3+(1-n_s)}\,,
\ea
where we used the notation introduced in \eqref{Legendreexp}. The explicit calculation of \cite{Endlich:2013jia} gives the leading order results
\begin{align}\label{b0}
b_{0}^{(0)}&=-\frac{10}{9}\frac{\cF_{Y}}{\cF \e c_{L}^{2}}+\dots& c_{1}^{(1)}&=\frac{20}{9}\frac{\cF_{Y}}{\cF \e c_{L}^{2}}+\dots\,,
\end{align}
where the dots represent $  \O(1) $ terms that were not calculated in \cite{Endlich:2013jia} because they are subleading in the $  \e\to 0 $ limit. We see that these results indeed obey our soft theorem \eqref{st1} to the appropriate order.


\subsection{Mixed scalar-tensor soft theorems}

The derivation of the mixed scalar-tensor consistency relation is very similar to the vector-tensor derivation above. First note that in the absence of long vector modes, the relevant OPE at leading order is
\ba
\zeta_{\bfk+\bfq/2}\zeta_{-\bfk+\bfq/2}&=&f^{ij}_T(\bfk)\gamma_{ij}(\bfq)+g^{ij}_T(\bfk)\dot{\gamma}_{ij}(\bfq)\\ \nn
&&+f^{ij}(\bfk)(\partial_i \partial_j \vpi_L)_{\bfq}+g^{ij}(\bfk)(\partial_i \partial_j \dot{\vpi}_L)_{\bfq} \, ,
\ea
where we anticipated the result and only wrote the terms that contribute to the soft theorem. The commutator and expectation value on the left hand side is unchanged \eqref{lhsmixed}, and for the right hand side the tensor part was given in \eqref{rhsmixed}. The scalar part in this case reads 
\be
[Q,(\partial_i \partial_j \vpi_L)_{\bfq}]=\sum_{a=1,2}\tilde{\omega}^a_{ij} {\cal I}_a(t) (2\pi)^3\delta^3(\bfq)\,.
\ee
Then, using the decomposition \eqref{sdecomp} and \eqref{mixeddecomp}, we find, analogously to the vector-tensor case, 
\ba
2f_T(k)+f_{\text{aniso}}(k)&=&-k\partial_{k}P_{\zeta}(k)\,,\\ \nn
2g_T(k)+g_{\text{aniso}}(k)&=&0\,.
\ea
The relevant correlators with the long modes can be written as
\ba
\la\gamma^s(-\bfq)\zeta_{\bfk+\bfq/2}\zeta_{-\bfk+\bfq/2}\ra'&=&\hat{k}^i\hat{k}^j\ep^s_{ij}(\hat{q})\,\left( f_T(k)\,P_{\gamma}(t,q)+\dfrac{1}{2}g_T(k) \dot{P}_{\gamma}(t,q)\right)\,,\\ \nn
\la \zeta_{-\bfq}\zeta_{\bfk+\bfq/2}\zeta_{-\bfk+\bfq/2}\ra'&=&(n_s-1)P_\zeta(k)\,P_\zeta (t,q)+ f_{\text{aniso}}(k)(3\cos^2\theta-1)\,P_\zeta(t,q) \,\\ \nn
&&+\dfrac{1}{2}g_{\text{aniso}}(k)(3\cos^2\theta-1) \dot{P}_\zeta(t,q)\,.
\ea
This OPE is also consistent with the quadrupole structure of the solid squeezed limit obtained by the background wave method in \cite{Endlich:2013jia}.
We again assume the existence of a decaying mode, namely
\be
\lim\limits_{q\to 0} \left(\dfrac{\dot{P}_{\gamma}(t,q)}{P_{\gamma}(t,q)}-\dfrac{\dot{P}_\zeta(t,q)}{P_\zeta(t,q)}\right)=0\,,
\ee
to arrive at the following consistency relation
\ba
\label{mixedST}
\boxed{2\,b^{(0)}_0(k)+\dfrac{1}{2}a_2^{(0)}(k)=3+(1-n_s)}\,,
\ea
again in the notation of \eqref{Legendreexp}. The coefficient $a_2^{(0)}$ in the set up of \cite{Endlich:2013jia} is found to be
\be
a_2^{(0)}(k)=+\dfrac{40}{9}\frac{\cF_{Y}}{\cF \e c_{L}^{2}}+..\,,
\ee
Together with \eqref{b0}, this agrees with the consistency relation we just derived up to $  \O(1) $  terms that were neglected in \cite{Endlich:2013jia}. The relation \eqref{mixedST} had already been noticed around Eq.~(6) of \cite{Bordin:2016ruc}. Here we have fleshed out their findings and performed two important additional checks. First, we proved that the relevant adiabatic mode does extend to non-zero momentum (finite wavelength), thus showing that the soft theorem holds for physical perturbations. Second, we found two adiabatic modes for the two possible time dependencies that such mixed scalar-tensor mode can exhibit. Had we had only one adiabatic mode then one time dependence would have been non-adiabatic. This would have implied that the soft theorem \eqref{fre} receives corrections proportional to the time derivative of the long mode, which is suppressed by slow-roll parameters but not by powers of the momenta. As our calculation shows this is not the case for the solid.

\subsection{Counter-collinear trispectrum}
A counter-collinear trispectrum \cite{Seery:2008ax} with equal legs can be easily obtained 
by squaring the OPE for $\zeta_{-\bfk-\bfq/2}\zeta_{+\bfk-\bfq/2}$, and one finds 
\ba
\nn
\lim_{q\to 0}\langle \zeta_{-\bfk-\bfq/2}\zeta_{+\bfk-\bfq/2}\zeta_{-\bfk+\bfq/2}\zeta_{+\bfk+\bfq/2}\rangle '&=& f_T^2(k)P_{\gamma}(q)-f^2_V(k)\,q^2P_V(q)\cos^2\theta\,(1-\cos^2\theta)\\ &&+9\,f^2_{\text{aniso}}(k)P_\zeta(q)\,(\cos^2\theta-1/3)^2\,+...\,,
\ea
where $\cos\theta=\hat{q}\cdot\hat{k}$, and dots stand for terms that are suppressed by slow-roll parameters in solid inflation, such as those proportional to the time derivative of the long mode power spectrum. This four-point function encodes the tree-level exchange of a soft scalar, vector and tensor. 

It is useful to expand this trispectrum in Legendre polynomials $P_\ell(\cos\theta)$ as 
\ba 
\lim_{q\to 0}\langle \zeta_{-\bfk-\bfq/2}\zeta_{+\bfk-\bfq/2}\zeta_{-\bfk+\bfq/2}\zeta_{+\bfk+\bfq/2}\rangle'=P_\zeta(k)^2\,P_\zeta(q)\sum d_\ell(k,q)\,P_{\ell}(\cos\theta)\,.
\ea
Then one finds that $d_1=d_3=d_{\ell\,>4}=0$, and
\ba
P_\zeta(k)^2\,d_0&=&\dfrac{P_\gamma(q)}{P_\zeta(q)} f^2_T(k)-\dfrac{2}{15}\dfrac{q^2P_V(q)}{P_\zeta(q)}f^2_V(k)+\dfrac{4}{5}f^2_{\text{aniso}}(k)\,,\\ \nn
P_\zeta(k)^2\, d_2&=&-\dfrac{2}{21}\dfrac{q^2P_V(q)}{P_\zeta(q)}f^2_V(k)+\dfrac{8}{7}f^2_{\text{aniso}}(k)\,,\\ \nn
P_\zeta(k)^2\, d_4&=&+\dfrac{8}{35}\dfrac{q^2P_V(q)}{P_\zeta(q)}f^2_V(k)+\dfrac{72}{35}f^2_{\text{aniso}}(k)\,.
\ea
Since $\frac{P_V(q)}{P_\zeta(q)}$ is not directly observable, we eliminate it among $d_0$, $d_2$ and $d_4$. Then, by making use of 
\be
f^2_{\text{aniso}}(k)= 4f_T^2(k)+{\cal O}(1)\,, 
\ee
we arrive\footnote{Note that the kinematics of \eqref{} enforces $d_0-d_2+d_4/6>0$, $5d_4-9d_2>0$ and $\frac{5}{12}d_4+d_2>0$. } at the following consistency relation:
\be
\label{tricc}
d_0-\left( 1+\frac{r}{16} \right)d_2+\left( \frac{1}{6}-\frac{r}{16} \right)d_4= \O(1)\,,\quad r=\frac{4P_\gamma(q)}{P_\zeta(q)}\,.
\ee
As opposed to the previous consistency conditions, \eqref{tricc} is less model-independent, as we have derived it in the set up of solid inflation, where ${\cal O}(1)$ and ${\cal O}(\epsilon)$ contributions are negligible and $d_0,d_2,d_4\gg 1$.


\subsection{Mixed ${\cal O}(q)$ scalar-tensor soft theorem}
\label{sample}
To show how our approach can capture the ${\cal O}(q)$ behavior of correlators, we study the the four point function or trispectrum with a soft scalar, i.e.~$\la \zeta_\bfq\zeta^3 \ra$, or a soft tensor, i.e.~$\la \gamma_\bfq\zeta^3\ra$. The reason for going beyond the bispectrum is that every ${\cal O}(q)$ correction to the bispectrum soft theorems vanishes by the parity arguments given around \eqref{parity}.

The OPE in this case can be written as
\ba
\nn
\zeta_{\bfk_1+\bfq/3}\zeta_{\bfk_2+\bfq/3}\zeta_{-\bfk_1-\bfk_2+\bfq/3}&=&{\cal O}(q^0)+{\cal F}^{ijl}(\bfk_a)\Big(\partial_i\partial_j\partial_l\phi_L\Big)_{\bfq}+{\cal F}_T^{ijl}(\bfk_a)\Big(\partial_l \gamma_{ij}\Big)_{\bfq}\\
&&\qquad +\cG^{ijl}(\bfk_a)\Big(\partial_i\partial_j\partial_l\dot{\phi}_L\Big)_{\bfq}+\cG_T^{ijl}(\bfk_a)\Big(\partial_l \dot{\gamma}_{ij}\Big)_{\bfq}
\ea
By means of the same adiabatic modes as in \eqref{tsmixed}, one could find consistency relations among ${\cal O}(q^0)$ coefficients as well. However, here we focus on the subleading ${\cal O}(q)$ part, to given an example of show how higher order adiabatic modes lead to relations among higher order coefficients in the soft limit.

It is useful to exploit the SO(3) symmetry to decompose the OPE coefficients, so we take
\ba
\nn
{\cal F}^{ijl}(\bfk_1,\bfk_2)&=&{\cal F}^a(k_1,k_2,k_3)\hat{k}_1^i\hat{k}_1^j\hat{k}_1^l+\dfrac{1}{3}{\cal F}^b(k_1,k_2,k_3)\Big(\hat{k}_1^i\hat{k}_2^j\hat{k}_2^l+\hat{k}_1^j\hat{k}_2^i\hat{k}_2^l+\hat{k}_1^l\hat{k}_2^i\hat{k}_2^j\Big)\\
&&+\dfrac{1}{3}{\cal F}^c(k_1,k_2,k_3)\Big(\hat{k}_1^i\delta_{jl}+\hat{k}_1^j \delta_{il}+\hat{k}_1^l\delta_{ij}\Big)+(1\leftrightarrow 2)\,,
\ea
where $k_3=|\bfk_1+\bfk_2|$, 
and similarly one can define $\cF^a_T, \cF_T^b, \cG^a,\cG^b,\cG^c, \cG_T^a$ and $\cG_T^b$. Notice that we will not need $\cF^c_T$ and $\cG^c_T$. Therefore, the OPE alone dictates the following structure
\ba
\label{sctri}
&&\dfrac{1}{P_\zeta(q)}\la \zeta_{-\bfq}\zeta_{\bfk_1+\bfq/3}\zeta_{\bfk_2+\bfq/3}\zeta_{-\bfk_1-\bfk_2+\bfq/3}\ra'=\\ \nn
&&\qquad {\cal O}(q^0)+3iq\Big[\,S_{\text{I}}(k_1,k_2,k_3)\cos^3\theta_1+S_{\text{II}}(k_1,k_2,k_3)\, \cos\theta_1\cos^2\theta_2+\\ \nn
&&\qquad\quad +S_{\text{III}}(k_1,k_2,k_3)\,\cos\theta_1+ (1\leftrightarrow 2)\Big]+{\cal O}(q^2)\,,
\ea
whereas a trispectrum with a soft graviton must take the form
\ba
\label{grtri}
&&\dfrac{1}{P_\gamma(q)}\la \gamma_{-\bfq}\zeta_{\bfk_1+\bfq/3}\zeta_{\bfk_2+\bfq/3}\zeta_{-\bfk_1-\bfk_2+\bfq/3}\ra'=\\ \nn
&& {\cal O}(q^0)+iq\Big[T_{\text{I}}(k_1,k_2,k_3)\hat{k}_1^i\hat{k}_1^j\epsilon_{ij}(\hat{q})\cos\theta_1\\ \nn
&&+T_{\text{II}}(k_1,k_2,k_3)(2\cos\theta_2\, \hat{k}_1^i\hat{k}_2^j\epsilon_{ij}(\hat{q})+\cos\theta_1 \hat{k}_2^i\hat{k}_2^j\epsilon_{ij}(\hat{q}))+(1\leftrightarrow 2)\Big]\,,
\ea
In these expressions we have introduced
\ba
S_{\text{I}}(k_1,k_2,k_3)&\equiv&\cF^a(k_1,k_2,k_3)+\dfrac{\dot{P}_\zeta(q)}{2P_\zeta(q)}\cG^a(k_1,k_2,k_3)\,,\\ \nn
S_{\text{II}}(k_1,k_2,k_3)&\equiv&\cF^b(k_1,k_2,k_3)+\dfrac{\dot{P}_\zeta(q)}{2P_\zeta(q)}\cG^b(k_1,k_2,k_3)\,,\\ \nn
S_{\text{III}}(k_1,k_2,k_3)&\equiv&\cF^c(k_1,k_2,k_3)+\dfrac{\dot{P}_\zeta(q)}{2P_\zeta(q)}\cG^c(k_1,k_2,k_3)\,,
\ea
and 
\ba
T_{\text{I}}&\equiv&\cF^a_T(k_1,k_2,k_3)+\dfrac{\dot{P}_\gamma(q)}{2P_\gamma(q)}\cG^a_T(k_1,k_2,k_3)\,,\\ \nn
T_{\text{II}}&\equiv&\dfrac{1}{3} \left[ \cF^b_T(k_1,k_2,k_3)+\dfrac{\dot{P}_\gamma(q)}{2P_\gamma(q)}\cG^b_T(k_1,k_2,k_3) \right] \,.
\ea
Acting with the symmetry associated with the gradient adiabatic modes in \eqref{gradient} induces the following transformations
\ba
[Q,(\partial_i \partial_j \vpi_L)_{\bfq}]_{\text{nlin}}=[Q,\gamma_{ij}(\bfq)]_{\text{nlin}}=&\sum_{a} i\,M^a_{ijk}\,\cI_a(t)\,(2\pi)^3\dfrac{\partial}{\partial q^k}\delta^3(\bfq)\,,\\ \nn
[Q,(\partial_i\partial_j\partial_k \vpi_L)_\bfq]_{\text{nlin}}=[Q,(\partial_k\gamma_{ij})_\bfq]_{\text{nlin}}=&\sum_{a}M^a_{ijk}\cI_a(t)(2\pi)^3\delta^3(\bfq)\,,\\ \nn
[Q,\zeta_\bfk]_{\text{lin}}=&\dfrac{1}{2}\sum_{a} \,M^a_{ijl}\,\cI_a(t)\,k^i\dfrac{\partial}{\partial k^j}\dfrac{\partial}{\partial k^l}\zeta_\bfk\,.
\ea
Thus
\ba
\nn
&&\la [Q,\zeta_{\bfk_1}\zeta_{\bfk_2}\zeta_{\bfk_3}]_{\text{lin}}\ra=\\ \nn
&&=\dfrac{1}{2}\sum_a\,M_{ijl}^a\,{\cal I}_a(t)\Big[\Big(\dfrac{k_1^i\,k_1^j}{k_1}\dfrac{\partial B}{\partial k_1}+ \dfrac{k_2^i\,k_2^j}{k_2}\dfrac{\partial B}{\partial k_2}+\dfrac{k_3^i\,k_3^j}{k_3}\dfrac{\partial B}{\partial k_3}\Big)\partial_l \delta^3(\bfk_1+\bfk_2+\bfk_3)\\ \nn
&&+\Big(\dfrac{k_1^ik_1^jk_1^l}{k_1^3}(-\dfrac{\partial B}{\partial k_1}+k_1\dfrac{\partial^2 B}{\partial k_1^2})+\dfrac{k_2^ik_2^jk_2^l}{k_2^3}(-\dfrac{\partial B}{\partial k_2}+k_2\dfrac{\partial^2 B}{\partial k_2^2})\\ 
&&\qquad\qquad\qquad \qquad+\dfrac{k_3^ik_3^jk_3^l}{k_3^3}(-\dfrac{\partial B}{\partial k_3}+k_3\dfrac{\partial^2 B}{\partial k_3^2})\Big)\delta^3(\bfk_1+\bfk_2+\bfk_3)\Big]\,,
\ea
where $B(k_1,k_2,k_3)$ is the bispectrum of the hard modes. Matching the transformation of both sides of the OPE reveals that
\ba
&&{\cal F}^a+\cF^a_T=\dfrac{1}{2}\Big(-\dfrac{\partial B}{\partial k_1}+k_1\dfrac{\partial^2 B}{\partial k_1^2}+\dfrac{k_1^3}{k_3^3}\dfrac{\partial B}{\partial k_3}-\dfrac{k_1^3}{k_3^2}\dfrac{\partial^2 B}{\partial k_3^2}\Big)\,,\\ \nn
&&\cG^a+\cG^b=0\,,\\ \nn
&&\cF^b+\cF^b_T=\dfrac{1}{2}\dfrac{k_1k_2^2}{k_3^3}(-\dfrac{\partial B}{\partial k_3}+k_3\dfrac{\partial^2 B}{\partial k_3^2})\,,\\ \nn
&&\cG^c+\cG^c_T=0\,, 
\ea
and therefore we arrive at the following observable consistency relations for the coefficients appearing in \eqref{sctri} and \eqref{grtri}, 
\be
S_{\text{I}}+T_{\text{I}}&=&\dfrac{1}{2}\Big(-\dfrac{\partial B}{\partial k_1}+k_1\dfrac{\partial^2 B}{\partial k_1^2}+\dfrac{k_1^3}{k_3^3}\dfrac{\partial B}{\partial k_3}-\dfrac{k_1^3}{k_3^2}\dfrac{\partial^2 B}{\partial k_3^2}\Big)\,,\\ \nn
S_{\text{II}}+T_{\text{II}}&=&\dfrac{1}{2}\dfrac{k_1k_2^2}{k_3^3}(-\dfrac{\partial B}{\partial k_3}+k_3\dfrac{\partial^2 B}{\partial k_3^2})\,.
\ee


\section{Conclusion}

A solid is characterized by the non-standard symmetry breaking pattern \eqref{symmpattern}. When applied to cosmology, this implies the appearance of anisotropic stresses, the non-conservation of $\zeta$ on superHubble scales and the violation of the celebrated Maldacena's consistency relations.

In this work we have shown how the usual construction of adiabatic modes is modified in the case of a solid. Using the same methods as \cite{Pajer:2017hmb}, we derive all solid adiabatic modes and discussed explicitly the leading ones. We re-derived the scalar soft theorem of \cite{Bordin:2017ozj}, stating that Maldacena's consistency relation is still valid upon angular averaging. We found also mixed scalar-tensor (already noticed in \cite{Bordin:2016ruc}) and vector-tensor consistency relations. All these relations rely on the symmetry breaking pattern only, not on the specific properties of the solid. Even though the standard construction in uniform-density gauge relies on residual large diffs, we showed that we can still construct adiabatic modes even in uniform-density unitary gauge, which does not possess any residual diffs.

Soft theorems studied for single clock cosmologies are model-independent: they reflect the way in which Poincare's symmetries are broken in these models, and not the details of individual single field scenarios.
Going even further, one might conjecture that soft theorems have the potential to probe the symmetry breaking pattern that led to cosmic inflation. Our work supports this conjecture, showing the distinct soft theorems emerging from the symmetry breaking pattern of a solid coupled to gravity. There are several avenues for future investigation:
\begin{itemize}
\item It would be interesting to find the soft theorems corresponding to yet other symmetry breaking patterns, such as Gauge-flation \cite{Maleknejad:2011jw}, gaugid inflation \cite{Piazza:2017bsd} and supersolid inflation \cite{Bartolo:2015qvr,Ricciardone:2016lym}. 
%
\item It would be nice to better understand the interplay between internal and space-time symmetries in the construction of adiabatic modes and the resulting soft theorems. 
\item Reheating can have a non-trivial impact on the predictions of solid inflation for superHubble perturbations in general, and for the soft theorems derived in this work. From a phenomenological perspective, it would be important to have a better understanding of these effects, before confronting solid soft theorems with observations. 	
\end{itemize}

 
\section*{Acknowledgements} It is a pleasure to thank Paolo Creminelli, Garrett Goon, Luca Santoni, and Mehrdad Mirbabayi for useful discussions.
D. vd W. and E. P. have been supported by the Delta-ITP consortium, a program of
the Netherlands Organization for Scientific Research (NWO) that is funded by the Dutch
Ministry of Education, Culture and Science (OCW). This work is part of the research programme
VIDI with Project No. 680-47-535, which is (partly) financed by the Netherlands
Organisation for Scientific Research (NWO). S.J. would like to thank the Utrecht University for the warm hospitality while a part of this research was in progress. S.J. would like to thank the Iranian National Elites Foundations (BMN) for their financial support during the completion of this work. 


\appendix


\section{From uniform density to uniform-density unitary gauge}\label{app:transformation}

consistency relations for $\zeta$ are gauge independent. In this appendix we show how this works out for the scalar consistency relations we found in uniform-density unitary gauge and in uniform density gauge. To relate the two results, we need to understand the relation between $A$ and $\zeta$ up to second order since we are considering three point functions. The second order relation between the two can be obtained as follows. 

In uniform-density unitary gauge, $A$ is defined through 
\be
2A=\dfrac{1}{a^2}\left(\delta g_{ii}-\nabla^{-2}\partial_i \partial_j \delta g_{ij}\right)\,.
\ee
To compute the metric in this gauge, but in terms of $\zeta$, we should apply a coordinate change from uniform density gauge, i.e.~
\begin{align}
x^{i}_{UDU}=x^{i}_{U}-\epsilon^{i}(x_{UD}) \, .
\end{align}
Then the metric in uniform-density unitary gauge up to second order becomes
\ba
\dfrac{1}{a^2}\delta g_{ij}&=&2\zeta \delta_{ij}-2\partial_{(i}\ep^{j)}+\partial_i \ep^k\partial_j \ep^k\\ \nn
&&-4\zeta \partial_{(i}\ep^{j)}+\delta_{ij}\left(2\zeta^2-2\ep^k\partial_k \zeta\right)\,.
\ea
Finally, we can relate $\epsilon^{i}$ to the field fluctuations from the definitions
\begin{align}
x^{i}+\varphi^{i}\equiv \phi^{i \prime}(x)=\phi^{i}[f^{-1}(x)]\, ; \quad f(x)\equiv x^{i}-\epsilon^{i}(x)\, .
\end{align}
Solving this to second order in $\epsilon$ yields $\varphi^{i}=\epsilon^{i}+\epsilon^{k}\partial_{k}\epsilon^{i}$, which can in turn be inverted to give
\begin{align}
\epsilon^{i}=\varphi^{i}-\varphi^{k}\partial_{k}\varphi^{i} + \mathcal{O}(\varphi^{3})\, .
\end{align}
The  uniform-density unitary gauge metric in terms of $\zeta$ and $\varphi^{i}$ is then given by
\ba
\dfrac{1}{a^2}\delta g_{ij}&=&\left(2\zeta+2\zeta^2-2\vpi^k\partial_k \zeta\right)\delta_{ij}-2\partial_i \partial_j \vpi_L\\ \nn
&&+3\partial_i \vpi^k\,\partial_j \vpi^k+2\vpi^k\partial_i \partial_j \vpi^k-4\zeta\partial_i \partial_j \vpi_L\, ,
\ea
where $\partial_{i}\varphi_{L}=\vpi^i$. The relation between $\zeta$ and $A$ is therefore given by
\ba\label{nonlinearA1}
2A&=&2(2\zeta+2\zeta^2-2\vpi^k\partial_k \zeta)+3\partial_i \vpi^k \partial_i \vpi^k-\dfrac{3}{\nabla^2}\partial_i \partial_j (\partial_i \vpi^k \partial_j \vpi^k)\\ \nn
&&+2\vpi^k \nabla^2 \vpi^k-\dfrac{2}{\nabla^2}\partial_i \partial_j (\vpi^k \partial_i \partial_j \vpi^k)-4\zeta \nabla^2\vpi_L+4\dfrac{1}{\nabla^2}\partial_i \partial_j (\zeta \partial_i \partial_j \vpi_L)\,.
\ea
where $\vpi^i$ should be algebraicly written in terms of $\zeta$ , i.e.~up to linear order 
\be
\vpi^i=\dfrac{3}{\nabla^2}\partial_i \zeta\,.
\ee
The inverse of \eqref{nonlinearA1} can be easily obtained and is given by
\ba\label{nonlinearA}
4\zeta &=& 2A+2A^2-\partial_k B\partial_k A-\dfrac{3}{4}\partial_i \partial_k B \partial_i \partial_k B+\dfrac{3}{4}\dfrac{1}{\nabla^2}\partial_i \partial_j (\partial_i \partial_k B \partial_j \partial_k B)\\ \nn
&&-\dfrac{1}{2}\partial_k B \nabla^2 \partial_k B+\dfrac{1}{2\nabla^2}\partial_i \partial_j (\partial_k B \partial_i \partial_j \partial_k B)+\dfrac{1}{\nabla^2}\partial_i \partial_j (A \partial_i\partial_j B)\,,
\ea
where 
\be
B=-\dfrac{3}{\nabla^2}A\,.
\ee


\subsection{Consistency relation}
We wish to compute 
\ba
\lim_{q\to 0}\,\int \dfrac{1}{2}d(\cos\theta)\,\langle \zeta_{\bfq}\zeta_{\bfk_1}\zeta_{\bfk_2}\rangle^{\prime}\,,
\ea
with $\theta$ being the angle between $\bfq$ and $\bfk_1$. 
We do this by plugging in \ref{nonlinearA} inside the $ \zeta$ bispectrum. The linear term in \ref{nonlinearA} does not give any contribution to the squeezed bispectrum of $\zeta$ because by the soft-$ A$ theorem, the $ A$ bispectrum vanishes in the squeezed limit, \eqref{zero}. The remaining terms come from inserting the nonlinear terms in \ref{nonlinearA}. Let us understand the kinematics of this expression. Since $q\ll k_1, k_2$, and since $P_{A}(p)\sim p^{-3}$, the leading terms come from inserting a second order term in place of $\zeta_{\bfk_1}$ or $\zeta_{\bfk_2}$, and taking the momentum of one of the fields to zero. In that case, we use the adiabatic mode solution for the long mode, which reads:
\ba
A_{l}=A_{l}(t) \, ;\quad B_{l}=-\dfrac{1}{2}A_l(t)\,\bfx^2.
\ea
Since one of the fields has to be of this type, we only have to consider terms with at most one spatial derivative per $\partial_i B$ field. The inverse Laplacians require some care, but they can always be moved past long-wavelength modes. Then we can write out all long-short ("ls") contributions to \ref{nonlinearA} where one of the fields is of the above form. Including the right combinatorial factors, we find,
\ba
4\zeta_s^{(ls)}=-2\,A_l x^k\partial_k A_s+3A_l A_s+\dfrac{3}{2} A_l\dfrac{1}{\nabla^2}\left(x^k\nabla^2 \partial_k A_s\right ).
\ea
The part in brackets in the third term can be rewritten as 
\begin{align}
x^{k}\nabla^{2}\partial_{k}A_{s}=\nabla^{2}\left(x^{k}\partial_{k}A_{s}\right)-2\partial_{j}x^{k}\partial_{k}\partial_{j}A_{s}=\nabla^{2}\left(x^{k}\partial_{k}A_{s}-2A_{s}\right)\,.
\end{align}
Altogether, we thus find that the quadratic contribution to $\zeta_{s}$ is
\ba
\zeta_s^{(2)}=-\dfrac{1}{8}A_l x^k\partial_k A_s=-\dfrac{1}{2}\zeta_l x^k\partial_k \zeta_s^{(1)} \, .
\ea
Plugging this into the bispectrum of $\zeta$  precisely generates the angle-averaged Maldacena consistency relation, namely \eqref{Malda}. 


\section{Fields, derivatives and the OPE}\label{app:derivatives}

To determine how many spatial derivatives are required in the OPE used in this paper, it is often enough to understand the symmetries of the theory. For instance, let us consider \eqref{charge0}, but now suppose we allow for a dependence on $\varphi_{L}$ without derivatives on the right hand side, i.e.
\begin{align}
\zeta_{\bf{k}-\frac{1}{2}\q}\zeta_{-\bf{k}-\frac{1}{2}\q}\xrightarrow{\q\to 0}P(k)(2\pi)^{3}\delta^{3}(\q)+f(k)\zeta_{-\q}+\tilde{f}(k)\varphi_{L}+\cdots \, .
\end{align} 
Then we can exploit the fact that $\varphi_{L}\to \varphi_{L}+c(t)$ is a symmetry of the theory. Namely, the charge corresponding to the symmetry, $Q_{s}$, only impacts $\varphi_{L}$, and commutes with $\zeta$. Acting with the charge on both sides of the OPE then immediately tells us tells us that $\tilde{f}(k)=0$. A similar shift symmetry exists for $\partial_i \phi_L$, thus we need at least two spatial derivatives acting on $\varphi_{L}$ for it to have a nonvanishing contribution to the OPE, as in \eqref{OPE}.

In general, any operator that the theory is symmetric under its shift can be excluded form the OPE. This typically fixes the number of spatial derivatives per field one needs to consider before applying any other symmetries.

\bibliographystyle{utphys}
\bibliography{refs}

\end{document}